\documentclass[reprint, aps, prd, fleqn, eqsecnum, balancelastpage, showpacs, superscriptaddress, nofootinbib]{revtex4-1}

\usepackage[utf8]{inputenc}
\usepackage[T1]{fontenc}

\usepackage{graphicx}	
\usepackage[hidelinks]{hyperref}   
\usepackage{amsmath}	
\usepackage{amssymb}	
\usepackage{bm}         
\usepackage{esdiff}     
\usepackage{isomath}    
\usepackage{natbib}     

\newsavebox{\foobox}
\newcommand{\slantbox}[2][0]{\mbox{%
    \sbox{\foobox}{#2}%
    \hskip\wd\foobox
    \pdfsave
    \pdfsetmatrix{1 0 #1 1}%
    \llap{\usebox{\foobox}}%
    \pdfrestore
}}
\newcommand\unslant[2][-.25]{\slantbox[#1]{$#2$}}

\newcommand{\uppartial}{\unslant\partial}
\newcommand{\upPartial}{\unslant\partial\kern-0.8pt}

\begin{document}

\title{QED effects are negligible for neutron-star spin-down}

\author{Paul Ripoche}
\email{pripoche@ens-paris-saclay.fr}
\altaffiliation{Department of Physics and Astronomy, University of British Columbia, 6224 Agricultural Road, Vancouver, BC V6T 1Z4, Canada}
\affiliation{École normale supérieure Paris-Saclay, 61 avenue du Président Wilson, 94235 Cachan Cedex, France}
\affiliation{Sorbonne Université - Faculté des Sciences et Ingénierie, 4 place Jussieu, 75005 Paris, France}

\author{Jeremy Heyl}
\email{heyl@phas.ubc.ca}
\affiliation{Department of Physics and Astronomy, University of British Columbia, 6224 Agricultural Road, Vancouver, British Columbia V6T 1Z4, Canada}

\date{\today}

\begin{abstract}
The energy loss of a rotationally powered pulsar is primarily carried away as electromagnetic radiation and a particle wind. Considering that the magnetic field strength of pulsars ranges from about $10^8$ to $10^{15}$~G, one could expect quantum electrodynamics (QED) to play a role in their spin-down, especially for strongly magnetized ones (magnetars). In fact several authors have argued that QED corrections will dominate the spin-down for slowly rotating stars. They called this effect quantum vacuum friction (QVF). However, QVF was originally derived using a problematic self-torque technique, which leads to a dramatic overestimation of this spin-down effect. Here, instead of using QVF, we explicitly calculate the energy loss from rotating neutron stars using the Poynting vector and a model for a particle wind, and we include the QED one-loop corrections. We express the excess emission as QED one-loop corrections to the radiative magnetic moment of a neutron star. We do find a small component of the spin-down luminosity that originates from the vacuum polarization. However, it never exceeds one percent of the classical magnetic dipole radiation in neutron stars for all physically interesting field strengths. Therefore, we find that the radiative corrections of QED are irrelevant in the energetics of neutron-star spin-down. 
\end{abstract}


\maketitle

\section{Introduction}

Neutron stars are the final stage of the evolution of stars with a mass between $M\approx 9~\text{M}_{\odot}$ and an upper mass still not determined precisely. These objects have a radius of about $R\approx 10~\text{km}$ and a mass of about $M\approx 1.4~\text{M}_{\odot}$. Neutron stars, like most of astrophysical objects, rotate, and thus have a rotational energy. This energy reservoir can account for the energy loss in a neutron star, and the spin-down that follows. The bulk of the energy extracted from the rotation of a neutron star is carried away partly as electromagnetic radiation, and partly as a wind, called a pulsar wind; that wind is composed of electrons, positrons and likely ions, pulled off from the surface of a pulsar. For the principal population of pulsars we get a magnetic field at the pole centered at around $B_\text{\tiny p} \approx 10^{12}~\text{G}$. Another interesting population is the one of magnetars, for which $B_\text{\tiny p} > 10^{14}~\text{G}$. Having in mind the critical magnetic field derived in quantum electrodynamics (QED) is $B_\text{\tiny QED} = \frac{m_{e}^2 c^3}{e \hbar} \approx 4.4 \times 10^{13}~\text{G}$, we could expect QED effects to play a role in the energy loss of neutron stars.

\citet{2008EL.....8269002D} argue that strongly magnetized neutron stars (magnetars) lose energy primarily through a process called quantum vacuum friction (QVF), in which the magnetized vacuum surrounding a neutron star spins it down. More recently, \citet{2016ApJ...823...97C,2016RAA....16....9X} and \citet{2012EL.....9849001D} have continued to argue that QVF dominates the energy loss of slowly rotating pulsars and especially  magnetars. Quantum vacuum friction is a phenomenon related to the fact that quantum vacuum can be regarded as a standard medium with its own energy density and electromagnetic properties. Thus, QVF can be seen as QED corrections to the radiation reaction torque in electromagnetism. We show that the authors have vastly overestimated the size of this effect, because they have used an inappropriate approximation for the structure of the magnetic field near to the surface of a neutron star and a problematic self-torque technique to calculate this effect. They assume that the dipole field is retarded even near to the star, but the retardation only develops in the radiation zone of a dipole. Furthermore, they estimate the torque exerted on the star by the induced magnetization surrounding it. The calculation of self-forces in electrodynamics has a long and subtle development starting with \citet{1902AnP...315..105A,1904AnP...319..236A}. In particular one must be careful in choosing which components of the field to include in the calculation and even then it often proves difficult to get reasonable results \cite{PhysRevD.73.065006}. Additionally, the electromagnetic angular momentum is often wrongly neglected in the conservation of the total angular momentum \cite{2018AmJPh..86..839B}; and the expression of the self-torque is somewhat more complicated, than in \cite{2008EL.....8269002D}, in the presence of charged particles \cite{2018AmJPh..86..839B,2009PhRvD..80b4031G}, i.e. a neutron star surrounded by a magnetosphere, which is a more realistic scenario.
As in \citet{2008EL.....8269002D}, we first calculate energy losses by modeling the rotating neutron star as a rotating magnetic dipole moment; we do not use the problematic self-torque technique leading to QVF, but rather calculate the energy flow using the Poynting vector. We then apply the QED one-loop corrections, in the weak-field approximation, to find the one-loop corrections to the dipole energy loss rate. We generalize these results to  the magnetic and electric field calculated by \citet{1955AnAp...18....1D}, for a rotating neutron star in vacuum. We find the same result in the weak-field limit as for a rotating magnetic dipole, and extend the calculation to the strong-field regime. We find that quantum vacuum renormalizes the magnetic moment of the star \cite{1997JPhA...30.6485H} only by a small amount for all reasonable magnetic field strengths. Although using a different method and a general relativistic description, the work of \citet{2016A&A...594A.112P,2016MNRAS.456.4455P} goes along with our results.

\citet{1969ApJ...157..869G} argued that rotating neutron stars have a dense magnetosphere; we therefore cannot ignore this more realistic model in our study. Thus, we derive the QED one-loop corrections to the energy loss of a pulsar, including the effect of the Goldreich and Julian magnetosphere in our calculations. We find that they are similar in magnitude to the vacuum case. We come to the conclusion that QED one-loop corrections remain negligible in the presence of a pulsar magnetosphere.
\\~\\\indent
\emph{All the results are in Gaussian units, unless mentioned otherwise.}

\section{Motivations and astrophysical background}

We first study a simple model for a neutron star, by considering it as a rotating classical magnetic dipole moment. In the sections that follow, we examine more realistic models for a neutron star, using the \citet{1955AnAp...18....1D} fields and the \citet{1969ApJ...157..869G} magnetosphere.

\subsection{A rotating magnetic dipole in vacuum}
\label{sec:rot_dipole}

We use a dipole approximation for a neutron star as an orthogonal rotator:
\begin{enumerate}
\item Solid rotation with an angular speed $\Omega$
\item Dipole magnetic field with a dipole magnetic moment $m$, such that $\hat{\mathbfit{m}} \bm{\cdot} \hat{\bm{\Omega}} = \cos(\alpha=\frac{\pi}{2})=0$
\item Neutron star in vacuum
\end{enumerate}
where, in a Cartesian coordinate system, the magnetic moment of the star is
$\mathbfit{m}(t)=\left\langle m, i m, 0 \right\rangle e^{i\Omega t}$ and the angular velocity vector is $\bm{\Omega}=\left\langle 0,0,\Omega \right\rangle$.

\citet{2008EL.....8269002D} assume that the magnetic field surrounding a neutron star takes the following form
\begin{equation} 
\mathbfit{B}(\mathbfit{r}, t) = \frac{3\mathbfit{n}[\mathbfit{m}(t-r/c) \bm{\cdot} \mathbfit{n}] - \mathbfit{m}(t-r/c)}{r^3},
\label{eq:dipole_magnetic_field}
\end{equation}
where $\mathbfit{n}=\mathbfit{r}/r$. They use this assumption that the field is retarded everywhere to calculate the self-torque on a neutron star. In fact, the magnetic field of an oscillating dipole is not retarded in the immediate vicinity of a dipole and has several components \cite{1998clel.book.....J}
\begin{equation}
\begin{split}
\mathbfit{H}(\mathbfit{r}, t)  =~& \frac{ 3\mathbfit{n}[\mathbfit{m}(t)\bm{\cdot} \mathbfit{n}] - \mathbfit{m}(t) }{r^3} \left ( 1 - i k r \right ) e^{ikr} \\
& + k^2 \left [ \mathbfit{n} \bm{\times} \mathbfit{m}(t) \right ]
\bm{\times} \mathbfit{n} \frac{e^{ikr}}{r},
\end{split}
\label{dipole_B_field}
\end{equation}
where $k=\frac{\Omega}{c}$, $\mathbfit{n}=\frac{\mathbfit{r}}{r}$ and all of the terms vary as $e^{i\Omega t}$.

In the near zone, where $kr \ll 1$, we have
\begin{equation}
    e^{-ikr} = 1 - ikr - \frac{(kr)^2}{2} + \mathcal{O}[i(kr)^3],
\end{equation}
so
\begin{equation}
    1 - ikr = e^{-ikr} + \frac{(kr)^2}{2} + \mathcal{O}[(ikr)^3].
\end{equation}
Then, if we focus on the first term, we can write Eq.~(\ref{dipole_B_field}) as
\small
\begin{equation}
\mathbfit{H}(\mathbfit{r}, t)  \underset{kr \ll 1}{=} \frac{ 3\mathbfit{n}[\mathbfit{m}(t)\bm{\cdot} \mathbfit{n}] - \mathbfit{m}(t) }{r^3} \left ( 1 + e^{ikr} \frac{(kr)^2}{2} \right )
\label{dipole_B_field_bis}
\end{equation}
\normalsize
We take the real part of the field \footnote{Whereas the interacting fields are sometimes given in complex representation, we always use the real part of the fields (the fields being derived within linear Maxwell theory), before using nonlinear Maxwell theory.} to allow a direct comparison with Eq.~(\ref{eq:dipole_magnetic_field}), and we can then write Eq.~(\ref{dipole_B_field}) as 
\begin{equation}
\begin{split}
    \mathbfit{H}(\mathbfit{r}, t) \underset{kr \ll 1}{=}~& \frac{ 3\mathbfit{n}[\mathbfit{m}(t)\bm{\cdot} \mathbfit{n}] - \mathbfit{m}(t) }{r^3} \\
    & + \frac{k^2}{2}\frac{ 3\mathbfit{n}[\mathbfit{m}(t-r/c)\bm{\cdot} \mathbfit{n}] - \mathbfit{m}(t-r/c)}{r} \\
    & + k^2 \frac{\left [ \mathbfit{n} \bm{\times} \mathbfit{m}(t-r/c) \right ]
\bm{\times} \mathbfit{n}}{r} .
\end{split}
\end{equation}
Therefore, in the near zone, the first near-field component (term in $1/r^3$) is indeed not retarded with respect to the rotation of a dipole contrary to what \citet{2008EL.....8269002D} assumed; the retardation only starts in the radiation zone.

\subsection{Energy flow for a rotating magnetic dipole in vacuum}

Furthermore, an oscillating dipole also has an electric displacement \cite{1998clel.book.....J}
\begin{equation}
\mathbfit{D}(\mathbfit{r}, t) =  - k^2 \left [ \mathbfit{n} \bm{\times} \mathbfit{m}(t) \right ]\frac{e^{ikr}}{r} \left ( 1 - \frac{1}{ikr} \right ).
\end{equation}

The cross product of the fields yields the energy flow \cite{1998clel.book.....J,2009EJPh...30..983K}
\begin{equation}
\mathbfit{S} = \frac{c}{4\pi} \mathbfit{E} \bm{\times} \mathbfit{H},
\label{Poynting}
\end{equation}
where $\mathbfit{E} = \mathbfit{D} - \mathbfit{P}$ (in Lorentz–Heaviside units), and $\mathbfit{S}$ is the Poynting vector. The quantity $\mathbfit{P}$ denotes the polarization. In our case this is the vacuum polarization of QED. We will initially neglect this term to get the classical radiated electromagnetic power $P_\text{\tiny 0}$,
\begin{equation}
P_\text{\tiny 0} = \oint_A \mathbfit{S}_\textbf{\tiny 0} \bm{\cdot} \mathbfit{n} \, \text{d}A = \frac{2}{3} \frac{\Omega^4 m^2}{c^3},
\label{eq:dipolepower}
\end{equation}
where $\text{d}A=r^2 \sin\theta \, \text{d}\theta \, \text{d}\phi$ is the infinitesimal element of surface in spherical coordinates; the surface integral is over any sphere centered on the location of a dipole. 

We now extend this well-known result to include the effects of vacuum polarization that we can quantify using the effective Lagrangian of QED to one-loop order \cite{1936ZPhy...98..714H,1997PhRvD..55.2449H}
\begin{equation} 
\mathcal{L}(I,K) = \mathcal{L}_\text{\tiny 0}(I) + \mathcal{L}_\text{\tiny 1}(I,K),
\end{equation} 
where $\mathcal{L}_\text{\tiny 0}$ is the linear Lagrangian and $\mathcal{L}_\text{\tiny 1}$ is the radiative corrections to the Lagrangian from QED. \citet{1936ZPhy...98..714H} derived that effective Lagrangian using electron-hole theory; \citet{1951PhRv...82..664S} later derived it using QED. The Lagrangian can be written in terms of the following Lorentz invariants \cite{1936ZPhy...98..714H}
\begin{equation} 
I = 2\left(\mathbfit{B}^2 - \mathbfit{E}^2\right),~
K = -\left(4\mathbfit{E} \bm{\cdot} \mathbfit{B}\right)^2, 
\end{equation}
such that 
\begin{equation}
\mathcal{L}_\text{\tiny 0}(I)= -\frac{1}{4} I.
\end{equation}
Although this Lagrangian $\mathcal{L}(I,K)$ was initially derived for an homogeneous field strength, it can also be used for slowly varying inhomogeneous fields. However, for those fields, the typical spatial scale of variation of inhomogeneities has to be much larger \cite{2015PhRvD..91h5027K} than the Compton wavelength of the electron, $\lambda_\text{\tiny C} \approx 2.4 \times 10^{-12}$ m. In our case the typical spatial scale at stake is of order of the radius of a neutron star, we can therefore use this Lagrangian.  

The polarization $\mathbfit{P}$ is given by \cite{Berestetskii}
\begin{equation}
\mathbfit{P} = \frac{\uppartial \mathcal{L}_\text{\tiny 1}}{\uppartial \mathbfit{E}},
\end{equation}
in Lorentz–Heaviside units. Specifically, we find that
\begin{equation}
\mathbfit{P} = -4 \mathbfit{E} \frac{\uppartial \mathcal{L}_\text{\tiny 1}}{\uppartial I} -32\mathbfit{B} \left (\mathbfit{E}\bm{\cdot} \mathbfit{B} \right )\frac{\uppartial \mathcal{L}_\text{\tiny 1}}{\uppartial K}.
\end{equation}
To lowest order in the radiative corrections (i.e. to first order in the fine-structure constant, $\alpha_\text{\tiny QED}=\frac{e^2}{\hbar c}\approx\frac{1}{137}$), we have $\mathbfit{B} \| \mathbfit{H}$. Therefore only the first term contributes; let us define
\begin{equation}
\mathbfit{S}_\textbf{\tiny 1} = -\frac{c}{4\pi} \mathbfit{P} \bm{\times} \mathbfit{H} = 4 \frac{\uppartial \mathcal{L}_\text{\tiny 1}}{\uppartial I} \mathbfit{S}_\textbf{\tiny 0},
\label{eq:S1}
\end{equation}
as the QED part of the Poynting vector, so that $\mathbfit{S} = \mathbfit{S}_\textbf{\tiny 0} + \mathbfit{S}_\textbf{\tiny 1}$. 

We could also perform this same calculation using the Minkowski form of the Poynting vector
\begin{equation}
\mathbfit{S} = \frac{c}{4\pi} \mathbfit{D} \bm{\times} \mathbfit{B},
\end{equation}
where $\mathbfit{B} = \mathbfit{H} + \mathbfit{M}$ (in Lorentz–Heaviside units). The quantity $\mathbfit{M}$ denotes the magnetization in our case of the vacuum, given by \cite{Berestetskii}
\begin{equation}
\mathbfit{M} = \frac{\uppartial \mathcal{L}_\text{\tiny 1}}{\uppartial \mathbfit{B}},
\label{magnetization}
\end{equation}
in Lorentz–Heaviside units. We get
\begin{equation}
\mathbfit{S} = \frac{c}{4\pi} \mathbfit{D} \bm{\times} \left (\mathbfit{H}+\mathbfit{M} \right )
= \frac{c}{4\pi} \mathbfit{D} \bm{\times} \left (\mathbfit{H}+ \frac{\uppartial \mathcal{L}_\text{\tiny 1}}{\uppartial \mathbfit{B}}\right ),
\end{equation}
where
\begin{equation}
\label{magnetization_field}
\mathbfit{B}=\mathbfit{H}+\frac{\uppartial \mathcal{L}_\text{\tiny 1}}{\uppartial \mathbfit{B}}
=  \mathbfit{H} + 4 \mathbfit{B} \frac{\uppartial \mathcal{L}_\text{\tiny 1}}{\uppartial I} - 32  \mathbfit{E} \left (\mathbfit{E}\bm{\cdot} \mathbfit{B} \right )\frac{\uppartial \mathcal{L}_\text{\tiny 1}}{\uppartial K},
\end{equation}
and 
\begin{equation}
\mathbfit{S}_\textbf{\tiny 1} = \frac{c}{4\pi} \mathbfit{D} \bm{\times} \mathbfit{M} 
=  4 \frac{\uppartial \mathcal{L}_\text{\tiny 1}}{\uppartial I} \mathbfit{S}_\textbf{\tiny 0},
\end{equation}
as before.

In the weak-field limit, the magnetic field strength $B$ at the surface of a neutron star is such that $B \ll B_\text{\tiny QED} \approx 4.4 \times 10^{13}~\mathrm{G}$. In such a regime, $\mathcal{L}_\text{\tiny 1}$  (to first order in K) is given by \cite{1936ZPhy...98..714H,1997PhRvD..55.2449H}
\begin{equation}
\mathcal{L}_\text{\tiny 1}(I,K) = \frac{\alpha_\text{\tiny QED}}{2\pi B_\text{\tiny QED}^2} \left( \frac{1}{180} I^2 - \frac{7}{720} K \right).
\end{equation}
We then find that 
\begin{equation}
\frac{\uppartial \mathcal{L}_\text{\tiny 1}}{\uppartial I} = \frac{\alpha_\text{\tiny QED}}{2\pi B_\text{\tiny QED}^2} \frac{I}{90}.
\label{eq:weakfielddlag}
\end{equation}

Using Eqs.~(\ref{eq:S1}) and (\ref{Poynting}), we get the additional QED radiated electromagnetic power $P_\text{\tiny 1}$,
\begin{equation}
P_\text{\tiny 1} = \oint_A \mathbfit{S}_\textbf{\tiny 1} \bm{\cdot} \mathbfit{n} \, \text{d}A = 
\frac{8\alpha_\text{\tiny QED}}{75 \pi}  \frac{m^2}{r^6  B_\text{\tiny QED}^2} \frac{2}{3} \frac{\Omega^4 m^2}{c^3}.
\label{eq:p1}
\end{equation}
This result is somehow a factor of $9$ bigger than in \cite{2016PhRvD..94d5021D}. An explanation might be found in the way \citet{2016PhRvD..94d5021D} derive the QED one-loop corrections, which might differ from ours.

If we take $r$ to be the radius of the star ($R$), we find that some additional electromagnetic energy is radiated through the surface to excite the polarization of the vacuum. Because Eq.~(\ref{eq:p1}) is valid in the weak-field limit, we can take $r$ to infinity and see that this additional radiative power vanishes as $r$ increases. As the vacuum has no energy sources or sinks, the total energy flux leaving the star must be conserved. To resolve this apparent paradox, we can assume that at infinity the total dipole moment of the star is somewhat larger than at the surface, due to the polarization of the vacuum (if we use the Abraham form of the Poynting vector), or the magnetization of the vacuum (if we use the Minkowski form). To account for this polarization, we use an expansion, to first order in $\alpha_\text{\tiny QED}$, of the magnetic moment $m$,
\begin{equation} 
m(r) = m_\text{\tiny 0} + m_\text{\tiny 1}(r),
\end{equation} 
where $m_\text{\tiny 0}$ is the bare magnetic dipole moment at the surface, and $m_\text{\tiny 1}(r)$ is an $r$-dependent correction to the magnetic moment, due to QED. $m_\text{\tiny 1}$ accounting for the conservation of the energy outside of the star and since we consider the neutron itself as a classical object (internal and crust effects are not part of our model), we set $m_\text{\tiny 1}(R) = 0$. We find
\small
\begin{equation}
m(r) = m_\text{\tiny 0}\left\{1 + \frac{\alpha_\text{\tiny QED}}{75 \pi}\left(\frac{2 m_\text{\tiny 0}}{R^3 B_\text{\tiny QED}}\right)^2\left[1-\left(\frac{R}{r}\right)^6\right] \right\}.
\end{equation}
\normalsize
Thus, the magnetic moment measured at infinity is slightly larger than at the surface of the star by an amount
\begin{equation}
m_\text{\tiny 1}(\infty) = \frac{4\alpha_\text{\tiny QED}}{75\pi} m_\text{\tiny 0} \frac{m_\text{\tiny 0}^2}{R^6 B_\text{\tiny QED}^2},
\label{eq:m1_dipole}
\end{equation}
where  $2m_\text{\tiny 0}/R^3 \ll B_\text{\tiny QED}$. 

\citet{1997JPhA...30.6475H} found a very similar expression for the radiative corrections to a static magnetic dipole of
\begin{equation}
m_\text{\tiny 1}(\infty) = \frac{4\alpha_\text{\tiny QED}}{135\pi} m_\text{\tiny 0} \frac{m_\text{\tiny 0}^2}{R^6 B_\text{\tiny QED}^2},
\end{equation}
in the weak-field limit, a factor of $9/5$ smaller than our expression. It is not surprising that we obtain the same scaling here as in \cite{1997JPhA...30.6475H} as both results are essentially angular averages of $\uppartial \mathcal{L}/\uppartial I$; however, in our case the average is weighted by a dipole radiation pattern (i.e. $\mathbfit{S}_\textbf{\tiny 0}$) and in the former case the weighting also includes an octopole term. 

Rather than treating the strong-field limit in the case of a simple rotating magnetic dipole, we examine, in the next sections, a more realistic field configuration for a rotating neutron star and examine both the weak-field and strong-field limits.

\subsection{The problematic QVF}
\subsubsection{Radiation reaction torque}
For the sake of our argumentation, we derive the radiation reaction torque (classical self-torque) in the $z$ direction, using the self-torque technique described in \cite{2008EL.....8269002D} and using Eq.~(\ref{eq:dipole_magnetic_field}). We however highlight erroneous assumptions made by \citet{2008EL.....8269002D}, and thus derive a more accurate estimate. 

The infinitesimal induced classical vacuum dipole moment, at a position $\mathbfit{r}$ is given by
\begin{equation}
\mathbfit{\textbf{d}m}(\mathbfit{r}, t) = \mathbfit{B}(\mathbfit{r}, t) \, \text{d}V,
\end{equation}
where $\text{d}V=r^2 \sin\theta \, \text{d}r \, \text{d}\theta \, \text{d}\phi$ is the infinitesimal element of volume in spherical coordinates. 

The infinitesimal induced classical vacuum dipole moment produces itself a retarded infinitesimal magnetic field at the center of the star, given by
\begin{equation}
\mathbfit{\textbf{d}B}({\bf 0}, t) = \frac{3\mathbfit{r}[\mathbfit{\textbf{d}m}\left(\mathbfit{r}, t-r/c \right) \bm{\cdot} \mathbfit{r}]}{r^5} - \frac{\mathbfit{\textbf{d}m}(\mathbfit{r}, t-r/c)}{r^3}.
\end{equation}

We then get the infinitesimal self-torque from the following formula \cite{1998clel.book.....J,Greiner}
\begin{equation}
\mathbfit{\textbf{d}\tau_\textbf{\tiny self}} = \mathbfit{m}(t) \bm{\times} \mathbfit{\textbf{d}B}({\bf 0}, t).
\label{inf_torque_class}
\end{equation}

In order to derive the classical self-torque, we integrate (\ref{inf_torque_class}) over the space outside of the star. However, unlike \citet{2008EL.....8269002D} who integrate directly from the surface of a neutron star, we assume that the field is retarded beyond a certain radius $u_\text{\tiny 0} R_\text{\tiny lc}$; we determine the cutoff scale, $u_\text{\tiny 0}$, below. The radius of the light cylinder of a neutron star, $R_\text{\tiny lc}$, is given by
\begin{equation} 
R_\text{\tiny lc} = \frac{c}{\Omega} = 4.8 \times 10^4 \left(\frac{P}{1~\text{s}}\right)~\text{km}, 
\end{equation}
where $P$ is the period of rotation of a neutron star. We get
\begin{equation}
\tau_\text{\tiny self} = \int_{r=u_\text{\tiny 0} R_\text{\tiny lc}}^{r=\infty}\int_{\theta=0}^{\theta=\pi}\int_{\phi=0}^{\phi=2\pi} (\mathbfit{\textbf{d}\tau_\textbf{\tiny self}} \bm{\cdot} \mathbfit{\hat{e}_z}) \, \text{d}V,
\label{torque_integral}
\end{equation}
The integration gives
\begin{equation}
\begin{split}
\tau_\text{\tiny self} =~& -\frac{8 \pi}{3} \frac{m^2 \Omega^3 \sin^2\alpha}{u_\text{\tiny 0}^3 c^3} \left[ 4 \text{Ci}(2 u_\text{\tiny 0})u_\text{\tiny 0}^3 + \cos(2 u_\text{\tiny 0})u_\text{\tiny 0} \right. \\
& \left. +~(1 - 2 u_\text{\tiny 0}^2) \sin(2 u_\text{\tiny 0}) \right],
\label{torque_classical_int}
\end{split}
\end{equation}
where $\text{Ci}$ is the cosine integral.

It is known \cite{1997AmJPh..65...81S} that for a uniformly rotating magnetic dipole, the expression of the radiation reaction torque has to agree with the one of the dipole torque. The latter is derived from classical electromagnetism \cite{Greiner}, and given by
\begin{equation}
	\tau_{\text{\tiny dipole}} = - \frac{2}{3 c^3}~m^2 \Omega^3 \sin^2(\alpha).
    \label{tauDipole}
\end{equation}

Therefore, equating those two torques sets the cutoff scale $u_\text{\tiny 0}$; we get
\begin{equation}
u_\text{\tiny 0}\approx 1.149.
\label{u_node}
\end{equation}

Consequently, the self-torque technique, and \textit{a fortiori} QVF, is only valid from around the radius of the light cylinder and not near the surface of a neutron star, as predicted by \citet{2008EL.....8269002D}.

Following the reasoning of \citet{2008EL.....8269002D}, $u_\text{\tiny 0}$ would be small and we would have
\begin{equation}
\tau_\text{\tiny self} \underset{u_\text{\tiny 0} \ll 1}{=} -8 \pi \frac{m^2 \Omega^3 \sin^2\alpha}{u_\text{\tiny 0}^2 c^3}.
\end{equation}
Consequently at the surface of a neutron star, $u_\text{\tiny 0} = \frac{\Omega R}{c}$, we would have
\begin{equation}
    \frac{\tau_\text{\tiny self}}{\tau_{\text{\tiny dipole}}} \underset{u_\text{\tiny 0} \ll 1}{=} 12\pi \left( \frac{R_\text{\tiny lc}}{R} \right)^2.
\end{equation}
As it will be demonstrated below, this scaling induces an overestimation of QVF by \citet{2008EL.....8269002D}.

\subsubsection{QVF in the weak-field limit}

We now consider the QED one-loop corrections to the magnetic field and we derive the additional self-torque from a QED-induced vacuum magnetization of the dipole field, following \cite{2008EL.....8269002D}.

Using Eq.~(\ref{eq:dipole_magnetic_field}), Eq.~(\ref{magnetization}) and Eq.~(\ref{eq:weakfielddlag}), the infinitesimal induced quantum vacuum dipole moment, at a position $\mathbfit{r}$ is given by
\begin{equation}
\mathbfit{\textbf{d}m}_\textbf{\tiny QVF}(\mathbfit{r}, t) = \frac{2\alpha_\text{\tiny QED}}{45\pi}\frac{\mathbfit{B}(\mathbfit{r}, t)^2} {B_\text{\tiny QED}^2} \mathbfit{B}(\mathbfit{r}, t) \, \text{d}V.
\end{equation}

The infinitesimal induced quantum vacuum dipole moment produces itself a retarded infinitesimal magnetic field at the center of the star, given by
\footnotesize
\begin{equation}
\mathbfit{\textbf{d}B_\textbf{\tiny QVF}}({\bf 0}, t) = \frac{3\mathbfit{r}[\mathbfit{\textbf{d}m}_\textbf{\tiny QVF}(\mathbfit{r},t-r/c) \bm{\cdot} \mathbfit{r}]}{r^5} - \frac{\mathbfit{\textbf{d}m}_\textbf{\tiny QVF}(\mathbfit{r},t-r/c)}{r^3}.
\end{equation}
\normalsize

We then get the infinitesimal self-torque,
\begin{equation}
\mathbfit{\textbf{d}\tau_\textbf{\tiny self, QVF}} = \mathbfit{m}(t) \bm{\times} \mathbfit{\textbf{d}B_\textbf{\tiny QVF}}({\bf 0}, t),
\label{inf_torque_qed}
\end{equation}
which leads to
\footnotesize
\begin{equation}
	\begin{split} 
    \tau_\text{\tiny self, QVF}=~& \frac{64}{212625}\frac{\alpha_\text{\tiny QED}} {B_\text{\tiny QED}^2 } \frac{m^4 \Omega^9 \sin^2(\alpha)}{u_\text{\tiny 0}^9c^9} \left \{ \left(\vphantom{\frac{1}{2}u_\text{\tiny 0}^8}  -2u_\text{\tiny 0}^8 + u_\text{\tiny 0}^6  \right. \right. \\
    & \left. \left. - 3 u_\text{\tiny 0}^4  + \frac{45}{2}u_\text{\tiny 0}^2 - 315 \right )\sin(2u_\text{\tiny 0})~+ \left [ \left (u_\text{\tiny 0}^6 - \frac{3}{2}u_\text{\tiny 0}^4 \right. \right. \right. \\
    & \left. \left. \left. + \frac{15}{2}u_\text{\tiny 0}^2 -\frac{315}{4} \right )\cos(2u_\text{\tiny 0}) + 4 \text{Ci}(2u_\text{\tiny 0})u_\text{\tiny 0}^8 \vphantom{\frac{1}{2}u_\text{\tiny 0}^8} \right] u_\text{\tiny 0} \right\}.
	\end {split}
    \label{torque_self_qed}
\end{equation}
\normalsize

Then, using the value of $u_\text{\tiny 0}$ from Eq.~(\ref{u_node}), we can evaluate (\ref{torque_self_qed}),
\begin{equation}
\tau_\text{\tiny self, QVF} = - 0.01397948990~\frac{\alpha_\text{\tiny QED} m^4 \Omega^9\sin^2(\alpha)} {c^9 B_\text{\tiny QED}^2 }.
\end{equation}
We note the dependence, here, on $\Omega^9$, which reduces the contribution of the magnetic field to the torque.

We then derive the following ratio
\begin{equation}
\frac{\tau_\text{\tiny self, QVF}}{\tau_{\text{\tiny dipole}}} = 0.02096923485~\frac{\alpha_\text{\tiny QED} m^2 \Omega^6}{c^6 B_\text{\tiny QED}^2},
\end{equation} 
and we get the following order of magnitude
\footnotesize
\begin{equation}
\frac{\tau_\text{\tiny self, QVF}}{\tau_{\text{\tiny dipole}}} = 1.7 \times 10^{-30} \left (\frac{B_0}{10^{12}~\text{G}}\right)^2 \left (\frac{R}{10~\text{km}}\right)^6 \left(\frac{P}{1~\text{s}}\right)^{-6}.
\end{equation}
\normalsize
We find, in the weak-field limit, that QVF is small compared to a classical dipole radiation. Consequently, QVF is negligible for neutron-star spin-down. 

Again, following the reasoning of \citet{2008EL.....8269002D}, $u_\text{\tiny 0}$ would be small and we would have
\begin{equation}
\tau_\text{\tiny self, QVF} \underset{u_\text{\tiny 0} \ll 1}{=} - \frac{16}{75}\frac{\alpha_\text{\tiny QED} m^4 \Omega^9\sin^2(\alpha)} {c^9 B_\text{\tiny QED}^2 u_\text{\tiny 0}^8},
\end{equation}
which would give the following ratio
\begin{equation}
\frac{\tau_\text{\tiny self, QVF}}{\tau_{\text{\tiny dipole}}} \underset{u_\text{\tiny 0} \ll 1}{=} \frac{8 \alpha_\text{\tiny QED}}{ 25 B_\text{\tiny QED}^2 } \frac{m^2 c^2} {\Omega^2 R^8},
\label{ratio_series_expansion}
\end{equation}
which explicitly depends on the radius of the star (not just the magnetic moment). 
\citet{2008EL.....8269002D} find the following value
\begin{equation}
  \frac{\tau_\text{\tiny self, QVF}}{\tau_{\text{\tiny dipole}}} \underset{\text{\tiny Dupays}}{=} \frac{9
    \alpha_\text{\tiny QED}}{128\pi B_\text{\tiny QED}^2} \frac{m^2 c^2}{R^8 \Omega^2},
\label{Dupays}  
\end{equation}
and \citet{2016ApJ...823...97C} get
\begin{equation}
  \frac{\tau_\text{\tiny self, QVF}}{\tau_{\text{\tiny dipole}}} \underset{\text{\tiny Coelho}}{=} \frac{2
    \alpha_\text{\tiny QED}}{25\pi B_\text{\tiny QED}^2} \frac{m^2 c^2}{R^8 \Omega^2}.
\label{Coelho}  
\end{equation}
Although the three results have different numerical coefficients, they have the same dependence on the dipole moment, spin frequency and stellar radius.

One can note the dependence on $\Omega^{-2}$ which supports the contribution of the field, hence the following overestimated order of magnitude
\footnotesize
\begin{equation}
\frac{\tau_\text{\tiny self, QVF}}{\tau_{\text{\tiny dipole}}} \underset{u_\text{\tiny 0} \ll 1}{=}
6.9 \left (\frac{B_\text{p}}{10^{12}~\text{G}}\right)^2 \left (\frac{R}{10~\text{km}}\right)^{-2} \left(\frac{P}{1~\text{s}}\right)^2.
\end{equation}
\normalsize

\subsubsection{Discussion}
The contribution of QVF, as also estimated by \citet{2008EL.....8269002D}, seems to become even more important for more slowly rotating neutron stars;  consequently, a realistic estimate of the near field and of the strong-field regime is crucial. Furthermore, QVF being only valid at around the radius of the light cylinder, a method taking into account the near field is needed to estimate the effects of QED on neutron-star spin-down. 

Thus, in the next section, we derive the QED one-loop corrections to the Poynting vector of a neutron star, using the Deutsch fields \cite{1955AnAp...18....1D}, instead of considering an additional spin-down effect such as QVF. 

\subsection{The Deutsch fields}
In his paper, \citet{1955AnAp...18....1D} idealizes a star as a sharply bounded, perfectly conducting sphere that rotates rigidly in vacuum. 

Let $\eta=\frac{\Omega r}{c}$ and $\delta=\frac{\Omega R}{c}$, $h_\text{\tiny 1}$ and $h_\text{\tiny 2}$ be spherical Bessel functions of the third kind (also known as spherical Hankel functions of the first kind) with argument $\eta$. Furthermore, primes will be used to denote derivatives with respect to the argument $h'_\text{\tiny 1}=\diff{h_\text{\tiny 1}}{\eta}$ and $h'_\text{\tiny 2}=\diff{h_\text{\tiny 2}}{\eta}$.
The expression $(~~~)_{\delta}$ denotes that the expression should be evaluated at the surface, i.e. $\eta \rightarrow \delta$. The general solution, for the external fields, derived by \citet{1955AnAp...18....1D} and corrected by \citet{1999PhR...318..227M}, is given here in Gaussian units, and with $r$, $\theta$ and $\phi$ the usual spherical coordinates.\\

\noindent\textit{\textbf{Deutsch magnetic field:}}
\footnotesize
\begin{equation} 
H_\text{r} = \frac{2m}{R^3} \left[\frac{R^3}{r^3} \cos\alpha \cos\theta + \frac{h_\text{\tiny 1}/\eta}{(h_\text{\tiny 1}/\eta)_\delta} \sin\alpha \sin\theta e^{i(\phi-\Omega t)} \right]
\label{Deutsch_magnetic_r}
\end{equation}
\begin{equation}
\begin{split}
H_\theta =~& \frac{m}{R^3} \left\{ \frac{R^3}{r^3} \cos\alpha \sin\theta + \left[ \left( \frac{\eta^2}{\eta h'_\text{\tiny 2} + h_\text{\tiny 2}}\right)_\delta h_\text{\tiny 2}~+ \right. \right. \\
& \left. \left. \left(\frac{\eta}{h_\text{\tiny 1}}\right)_\delta \left(h'_\text{\tiny 1} + \frac{h_\text{\tiny 1}}{\eta}\right) \vphantom{\frac{\eta^2}{\eta h'_\text{\tiny 2} + h_\text{\tiny 2}}} \right] \sin\alpha \cos\theta e^{i(\phi-\Omega t)} \right\}
\label{Deutsch_magnetic_theta}
\end{split}
\end{equation}
\begin{equation}
\begin{split}
H_\phi =~& \frac{m}{R^3} \left[ \left( \frac{\eta^2}{\eta h'_\text{\tiny 2} + h_\text{\tiny 2}}\right)_\delta h_\text{\tiny 2}\cos2\theta~+ \right. \\
& \left. \left(\frac{\eta}{h_\text{\tiny 1}}\right)_\delta \left(h'_\text{\tiny 1} + \frac{h_\text{\tiny 1}}{\eta}\right) \vphantom{\frac{\eta^2}{\eta h'_\text{\tiny 2} + h_\text{\tiny 2}}} \right] i\sin\alpha e^{i(\phi-\Omega t)}
\label{Deutsch_magnetic_phi}
\end{split}
\end{equation}\\
\normalsize

\noindent\textit{\textbf{Deutsch electric field:}}
\small
\begin{equation}
\begin{split}
E_\text{r} =~& \frac{\Omega R}{c}\frac{m}{R^3} \left[ \vphantom{\frac{\eta}{\eta h'_\text{\tiny 2} + h_\text{\tiny 2}}} -\frac{1}{2} \frac{R^4}{r^4} \cos\alpha (3\cos2\theta + 1) ~+ \right. \\
& \left. 3\left( \frac{\eta}{\eta h'_\text{\tiny 2} + h_\text{\tiny 2}}\right)_\delta\frac{h_\text{\tiny 2}}{\eta} \sin\alpha \sin2\theta e^{i(\phi-\Omega t)} \right]
\end{split}
\end{equation}
\begin{equation} 
\begin{split}
E_\theta =~& \frac{\Omega R}{c}\frac{m}{R^3} \left\{ -\frac{R^4}{r^4} \cos\alpha \sin 2\theta + \left[ \left( \frac{\eta}{\eta h'_\text{\tiny 2} + h_\text{\tiny 2}}\right)_\delta \right. \right. \\
& \left. \left. \frac{\eta h'_\text{\tiny 2} + h_\text{\tiny 2}}{\eta}\cos 2\theta - \frac{h_\text{\tiny 1}}{(h_\text{\tiny 1})_\delta} \vphantom{\frac{\eta}{\eta h'_\text{\tiny 2} + h_\text{\tiny 2}}} \right] \sin\alpha e^{i(\phi-\Omega t)} \right\}
\end{split}
\end{equation}
\begin{equation}
\begin{split}
E_\phi =~& \frac{\Omega R}{c}\frac{m}{R^3} \left[ \left( \frac{\eta}{\eta h'_\text{\tiny 2} + h_\text{\tiny 2}}\right)_\delta\frac{\eta h'_\text{\tiny 2} + h_\text{\tiny 2}}{\eta}~- \right. \\
& \left. \frac{h_\text{\tiny 1}}{(h_\text{\tiny 1})_\delta} \vphantom{\frac{\eta}{\eta h'_\text{\tiny 2} + h_\text{\tiny 2}}} \right] i\sin\alpha \cos \theta e^{i(\phi-\Omega t)}.
\label{Deutsch_electric_phi}
\end{split}
\end{equation}\\
\normalsize

Because $R \approx 10~\text{km}$, a useful approximation is to use the expressions for the fields when $R/R_\text{\tiny lc} \ll 1$.

Using \textsc{Maple} and some final calculations by hand, we derive the vectorial expression of the Deutsch magnetic field when $R \ll R_\text{\tiny lc}$
\footnotesize
\begin{equation} 
	\begin{split} 
\mathbfit{H}_\text{\textbf{\tiny D}}(\mathbfit{r}, t+r/c) \underset{\frac{\Omega R}{c}\rightarrow 0}{=}~& \frac{3\mathbfit{r}(\mathbfit{m}(t) \bm{\cdot} \mathbfit{r})}{r^5} - \frac{\mathbfit{m}(t)}{r^3} - \frac{3\mathbfit{r}[(\mathbfit{m}(t) \bm{\times} \bm{\Omega}) \bm{\cdot} \mathbfit{r}]}{c r^4} \\
& + \frac{\mathbfit{m}(t) \bm{\times} \bm{\Omega}}{c r^2} + \frac{\mathbfit{r}(\mathbfit{m}(t) \bm{\cdot} \bm{\Omega})(\mathbfit{r} \bm{\cdot} \bm{\Omega})}{c^2 r^3} - \\
& \frac{\Omega^2 \mathbfit{r} (\mathbfit{m}(t) \bm{\cdot} \mathbfit{r})}{c^2 r^3} - \frac{(\mathbfit{m}(t) \bm{\times} \bm{\Omega}) \bm{\times} \bm{\Omega}}{c^2 r},
	\end {split}
    \label{deutsch_magnetic_vectorial_approx}
\end{equation}
\normalsize
where, in a Cartesian coordinate system, we have
\footnotesize
\[
\mathbfit{r}=
\begin{pmatrix}
r\sin(\theta)\cos(\phi) \\
r\sin(\theta)\sin(\phi) \\
r\cos(\theta)
\end{pmatrix}
;
\mathbfit{m}(t)=
\begin{pmatrix}
m\sin(\alpha)\cos(\Omega t) \\
m\sin(\alpha)\sin(\Omega t) \\
m\cos(\alpha)
\end{pmatrix}
;
\bm{\Omega}=
\begin{pmatrix}
0 \\
0 \\
\Omega
\end{pmatrix}.
\]
\normalsize

One important thing to notice is that the Deutsch magnetic field intrinsically contains a retardation $t-r/c$, in the expression of the magnetic moment, no matter where the field is located in the space $r \geqslant R$. This retardation becomes explicit in Eq.~(\ref{deutsch_magnetic_vectorial_approx}), but it is already present in the spherical field components, only made implicit by the complex notation. Thus, we find here, in the first two terms of Eq.~(\ref{deutsch_magnetic_vectorial_approx}), the magnetic field of a classical magnetic dipole, but this time retarded, even near the surface of a neutron star. However, the two subsequent terms cancel the retardation for small values of $r$ as in the case of the rotating dipole in Sec.~\ref{sec:rot_dipole}. 

We also derived the full vectorial expression of the Deutsch magnetic field, using the decomposition in Eq.~(\ref{decomposition}) (see the appendix). One can note, in the full vectorial expression, that the intrinsic retardation of the field is now $t + R/c - r/c$. That comes from the continuity of the field at the surface of a neutron star, between the internal field and the external one. Therefore, the retardation is diminished further by the surface boundary conditions, by a factor of $R/c$, in comparison to the case of an usual rotating magnetic dipole.

\section{Energy flow in the Deutsch fields}

In the \citet{1955AnAp...18....1D} model, a neutron star essentially loses its energy in the form of electromagnetic radiation, so the Poynting vector quantifies the losses through the surface of the star. We first calculate the Poynting vector in a classical way, and then we study the QED one-loop corrections that can be applied to the macroscopic fields, and thus derive an additional Poynting vector from QED-induced vacuum polarization of the dipole field. We will go even further, using the conservation of the energy as a motivation to derive QED one-loop corrections to the magnetic dipole moment of a neutron star, to first order in $\alpha_\text{\tiny QED}$.
\\~\\
\emph{All the results using the Deutsch fields are indexed with a D in this chapter.}

\begin{figure*}
\includegraphics[width=14cm, trim={0 0 0 1cm},clip]{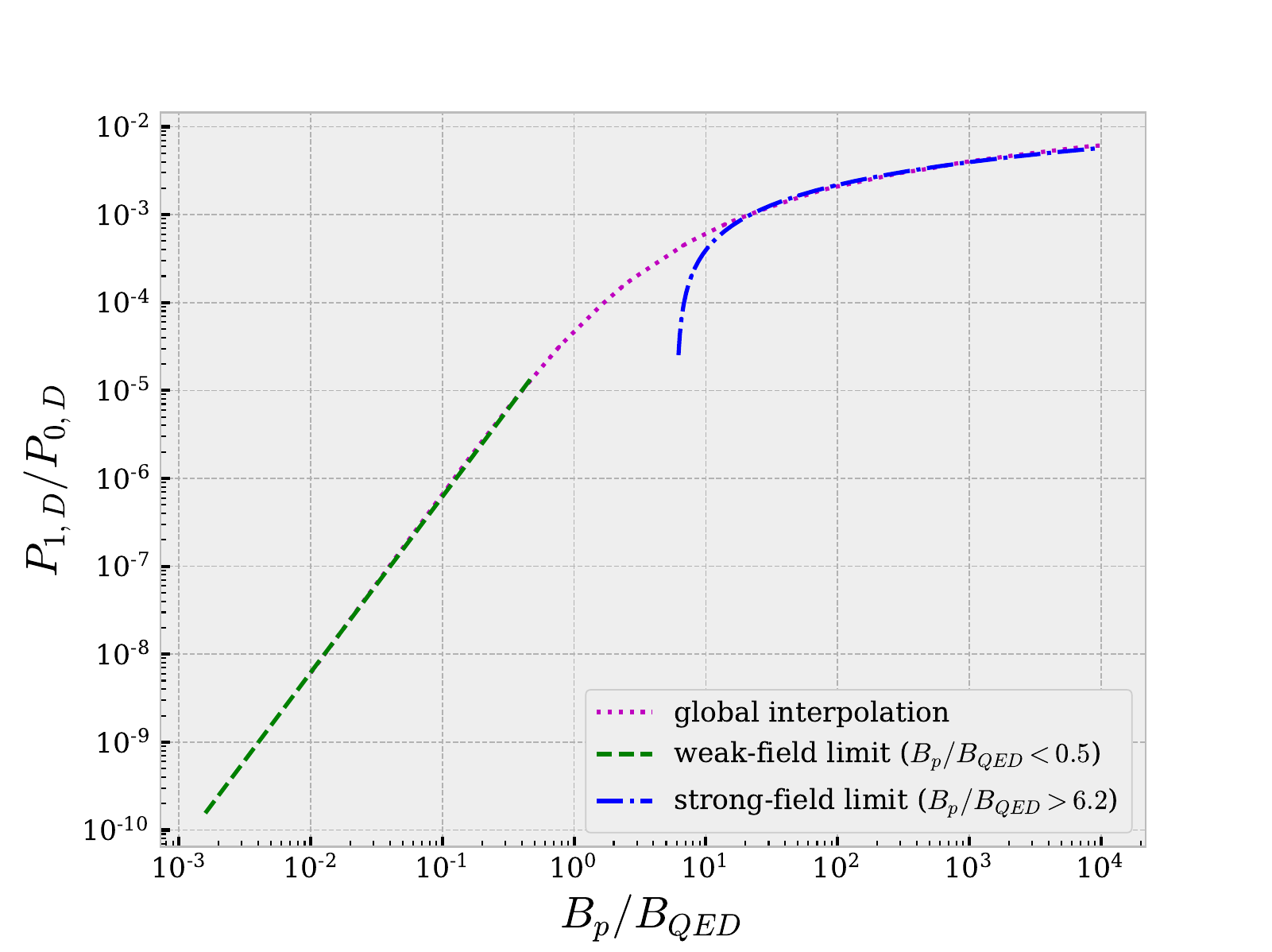}
\caption{The additional radiated power induced by QED for the Deutsch field, ($P_\text{\tiny 1,D}/P_\text{\tiny 0,D})$, as a function of the surface magnetic field ($B_\text{\tiny p}$), in the weak-field limit (green dashed curve), strong-field limit (blue dot-dashed curve) and a global interpolation (magenta dotted curve) using the results of \cite{1997JPhA...30.6475H}.}
\label{fig:delta_new}
\end{figure*}

\subsection{Classical approach}
As with the rotating dipole we calculate the Poynting vector 
\begin{equation} 
\mathbfit{S}_\text{\textbf{\tiny 0,D}} = \frac{c}{4\pi}  \mathbfit{E}_\text{\textbf{\tiny D}} \bm{\times} \mathbfit{B}_\text{\textbf{\tiny D}},
\end{equation}
and we integrate over a sphere centered on the star to get
\begin{equation}
P_\text{\tiny 0,D} = \oint_A \mathbfit{S}_\text{\textbf{\tiny 0,D}} \bm{\cdot} \mathbfit{n} \, \text{d}A
\label{int_em_D}.
\end{equation}

We are therefore only interested here in the radial component $S_\text{r,0,\tiny{D}}$ of the Poynting vector, with $r$, $\theta$ and $\phi$ the usual spherical coordinates,
\begin{equation} 
S_\text{r,0,\tiny{D}} = \frac{c}{4\pi} (E_{\theta,\text{\tiny D}}  B_{\phi,\text{\tiny D}} - E_{\phi,\text{\tiny D}} B_{\theta,\text{\tiny D}}).
\label{S_r}
\end{equation}

We get the radiated electromagnetic power by integrating over a surface $\text{d}A$
\footnotesize
\begin{equation}
 P_\text{\tiny 0,D}=
 \frac{2}{3} \frac{m^2 \Omega^4 \sin^2\alpha}{c \left(c^2+\Omega^2R^2\right)}
\frac{ 180c^6 - 12\Omega^4R^4c^2 + 8\Omega^6R^6}{ 180c^6 - 15 \Omega^4R^4c^2 + 5\Omega^6R^6},
\label{energy_em_D}
\end{equation}
\normalsize
which reduces to, assuming $R \ll R_\text{\tiny lc}$,
\begin{equation}
    P_\text{\tiny 0,D} \underset{R \ll R_\text{\tiny lc}}{=} \frac{2}{3c^3}m^2 \Omega^4 \sin^2\alpha.
    \label{energy_em_D_appr}
\end{equation}

We find back in Eq.~(\ref{energy_em_D_appr}) the radiated power derived in Eq.~(\ref{eq:dipolepower}) for an orthogonal magnetic dipole moment ($\alpha = \pi/2$), and the value calculated by \citet{1955AnAp...18....1D}. Equation~(\ref{energy_em_D}) is also important since it shows the dependence of the radiated power on the radius $R$ of the star as a correction to the dipole formula. Furthermore, as in Eq.~(\ref{eq:dipolepower}), the radiated power does not depend on the distance from the star ($r$).

\subsection{QED One-Loop corrections in the weak-field limit}
We  now work with a neutron star surrounded by quantum vacuum. As described in Sec.~\ref{sec:rot_dipole}, nonlinearities in the equations of the electromagnetic fields are introduced by the one-loop corrections of quantum electrodynamics. From Eq.~(\ref{eq:S1}), we get
\begin{equation}
\mathbfit{S}_\text{\textbf{\tiny 1,D}} =
 4 \frac{\uppartial \mathcal{L}_\text{\tiny 1}}{\uppartial I} \mathbfit{S}_\text{\textbf{\tiny 0,D}},
\label{eq:S1D}
\end{equation}
where we are only interested in the radial component as given by Eq.~(\ref{S_r}). We again use the weak-field limit from Eq.~(\ref{eq:weakfielddlag}), and integrate over a sphere of radius $r$ surrounding the star to get
\footnotesize
\begin{equation} 
\begin{split}
P_\text{\tiny 1,D}(r) =~&
\frac{2}{4725} \left[ 168 - 4\sin^2 \alpha \frac{\Omega^6 R^4 r^2}{c^6} - 20 \frac{\Omega^4 R^4}{c^4} \sin^2\alpha - \right. \\ 
& \left. - \left(  48 + 12\cos^2 \alpha \right ) \frac{\Omega^2R^4}{r^2 c^2}\right ] \frac{\alpha_\text{\tiny QED}}{\pi B_\text{\tiny QED}^2}\frac{\Omega^4 m^4\sin^2\alpha}{ c^3 r^6}.
\label{power_deu_qed}
\end{split}
\end{equation}
\normalsize
We find that the radiated power depends on the distance $r$ 
as with the rotating dipole. If we examine the limit where $R\ll R_\text{\tiny lc}$, we obtain 
\begin{equation} 
P_\text{\tiny 1,D}(r) \underset{R \ll R_\text{\tiny lc}}{=} \frac{8\alpha_\text{\tiny QED}}{75 \pi}  \frac{m^2}{r^6  B_\text{\tiny QED}^2} \frac{2}{3} \frac{m^2 \Omega^4 \sin^2\alpha}{c^3}.
\label{power_weak_limit}
\end{equation}

As in Sec.~\ref{sec:rot_dipole}. we can then derive the one-loop corrections to the magnetic dipole moment of the Deutsch field
\begin{equation}
m_\text{\tiny 1,D}(\infty) \underset{R \ll R_\text{\tiny lc}}{=} \frac{4\alpha_\text{\tiny QED}}{75\pi} m_\text{\tiny 0} \frac{m_\text{\tiny 0}^2}{R^6 B_\text{\tiny QED}^2},
\label{eq:m1_deutsch}
\end{equation}
which in the limit of $R\ll R_\text{\tiny lc}$ is identical to the results for the rotating dipole, Eq.~(\ref{eq:m1_dipole}).

Since Eq.~(\ref{power_weak_limit}) is true for each value of $r \geqslant R$, we can get an estimation of the energy loss rate at the surface of the star
\begin{equation}
\label{weak_ratio}
\frac{P_\text{\tiny 1,D}(R)}{P_\text{\tiny 0,D}} \underset{R \ll R_\text{\tiny lc}}{=} \frac{2\alpha_\text{\tiny QED}}{75\pi}\left(\frac{B_\text{\tiny p}}{B_\text{\tiny QED}}\right)^2,
\end{equation}
where
\begin{equation}
B_\text{\tiny p}=\frac{2m_\text{\tiny 0}}{R^3}
\end{equation}
is the magnetic field strength at the magnetic pole of a neutron star ($\theta =0$, $\phi=0$, $r=R$, and $\alpha=0$).

We can now evaluate the ratio of the additional spin-down power from vacuum polarization to the classical spin-down power to find
\begin{equation}
\frac{P_\text{\tiny 1,D}(R)}{P_\text{\tiny 0,D}} \underset{R \ll R_\text{\tiny lc}}{=} 3.2 \times 10^{-8} \left ( \frac{B_\text{\tiny p}}{10^{12}~\text{G}} \right )^2.
\end{equation}
We find for stars in the weak-field limit that the vacuum polarization contribution to the spin-down is negligible. However, it appears to increase as the square of the surface magnetic field, so perhaps it could be important for magnetars, therefore we must repeat the calculation in the strong-field limit.

\subsection{QED One-Loop corrections in the strong-field limit}

We now consider the case of magnetars, that is to say we use the QED one-loop corrections to the Deutsch field in the strong-field limit ($B_\text{\tiny p} \gg B_\text{\tiny QED}$). \citet{1997PhRvD..55.2449H}, as well as \citet{1976JETP...42..774R} and \citet{1976JPhA....9.1171D}, found the effective Lagrangian in the limit where $K$ is small (this is equivalent to $R B_\text{\tiny p}\ll R_\text{\tiny lc} B_\text{\tiny QED}$) which is generally true for the observed magnetars. We have, to the leading order,
\footnotesize
\begin{equation}
\mathcal{L}_\text{\tiny 1}(I,0) = \frac{\alpha_\text{\tiny QED}}{4\pi} I 
\left [ \frac{1}{6} \ln \left ( \frac{2I}{B_\text{\tiny QED}^2} \right ) - \frac{1}{3} + 4\zeta^{(1)}(-1) \right ],
\end{equation}
\normalsize
where $\zeta^{(1)}(-1)=-0.1654211437$ is the first derivative of the Riemann Zeta function evaluated in $-1$.

We then find
\footnotesize
\begin{equation}
\frac{\uppartial \mathcal{L}_\text{\tiny 1}}{\uppartial I}(I,0) = \frac{\alpha_\text{\tiny QED}}{4\pi} \left [ \frac{1}{6}  \ln \left ( \frac{2I}{B_\text{\tiny QED}^2} \right ) - \frac{1}{6}  + 4\zeta^{(1)}(-1) \right ].
\end{equation}
\normalsize
This expression is nearly constant over the surface of the star, since the strong-field regime is only valid until a radius $r_\text{\tiny s}$, not much bigger than $R$; further than that radius, the field switches to the weak-field regime. Consequently, within $r_\text{\tiny s}$, we take the field to vary slowly from the surface as
\begin{equation}
    I = 2\left( B_\text{\tiny p}\frac{R^3}{r^3} \right)^2.
\end{equation}

Proceeding as for the weak-field limit, we get the following radiated power
\footnotesize
\begin{equation}
\begin{split}
    P_\text{\tiny 1,D}(r) \underset{R \ll R_\text{\tiny lc}}{=}~& \frac{2\alpha_\text{\tiny QED}}{9 \pi} \frac{m^2 \Omega^4 \sin^2\alpha}{c^3} \left [ \ln \left ( \frac{B_\text{\tiny p}R^3}{B_\text{\tiny QED}r^3} \right ) \right. \\
    & \left. \vphantom{\ln \left ( \frac{B_\text{\tiny p}R^3}{B_\text{\tiny QED}r^3} \right )} + \ln(2) - \frac{1}{2}  + 12\zeta^{(1)}(-1) \right ].
\end{split}
\label{strong_power_deutsch}
\end{equation}
\normalsize

We can now derive the expression of $r_\text{\tiny s}$, given by $P_\text{\tiny 1,D}(r) = 0$,
\footnotesize
\begin{equation}
    r_\text{\tiny s} = R \left ( \frac{B_\text{\tiny p}}{B_\text{\tiny QED}}\right)^{\frac{1}{3}}\exp \left [-\ln(2) + \frac{1}{2} -12\zeta^{(1)}(-1) \right]^{-\frac{1}{3}};
\end{equation}
\normalsize
for example, for $B_\text{\tiny p} = 100 B_\text{\tiny QED}$, we get $r_\text{\tiny s} \approx 2.6 R$. 

Furthermore, the QED corrections,  within this radius, to the magnetic moment of a neutron star are purely geometric,
\begin{equation}
    m_\text{\tiny 1,D}(r \leq r_\text{\tiny s}) = \frac{1}{2\pi}\ln{\frac{r}{R}}.
\end{equation}

Then, at the surface of a neutron star, we have
\footnotesize
\begin{equation}
\frac{P_\text{\tiny 1,D}(R)}{P_\text{\tiny 0,D}} = \frac{\alpha_\text{\tiny QED}}{3\pi} \left [ \ln \left ( \frac{B_\text{\tiny p}}{B_\text{\tiny QED}} \right ) + \ln(2)- \frac{1}{2} + 12\zeta^{(1)}(-1)  \right ].
\end{equation}
\normalsize

We can now evaluate the ratio of the additional spin-down power from vacuum polarization to the classical spin-down power to find, in the strong-field limit,
\begin{equation}
\frac{P_\text{\tiny 1,D}(R)}{P_\text{\tiny 0,D}} = 7.7 \times 10^{-4} \ln \left (\frac{B_\text{\tiny p}}{2.6 \times 10^{14}~\text{G}}\right ).
\end{equation}

Again we will consider that the magnetic moment measured at infinity is slightly larger than at the surface of the star. However, since we consider distances further than $r_\text{\tiny s}$, we use the weak-field limit results, to yield
\footnotesize
\begin{equation}
\begin{split}
m_\text{\tiny 1,D}(r>r_\text{\tiny s}) =~& \frac{\alpha_\text{\tiny QED}}{6\pi} m_\text{\tiny 0} \left [ \ln \left ( \frac{2m_\text{\tiny 0}}{R^3 B_\text{\tiny QED}} \right ) + \ln(2) \right. \\
& \left. - \frac{1}{2} + 12  \zeta^{(1)}(-1) - \frac{8}{25}\frac{m_\text{\tiny 0}^2}{r^6 B_\text{\tiny QED}^2} \right ],
\label{eq:m1_deutsch_strong}
\end{split}
\end{equation}
\normalsize
where $2m_\text{\tiny 0}/R^3 \gg B_\text{\tiny QED}$. At infinity, we have 
\begin{equation}
\begin{split}
m_\text{\tiny 1,D}(\infty) =~& \frac{\alpha_\text{\tiny QED}}{6\pi} m_\text{\tiny 0} \left [ \ln \left ( \frac{2m_\text{\tiny 0}}{R^3 B_\text{\tiny QED}} \right ) \right. \\ 
& \left. \vphantom{\frac{2m_\text{\tiny 0}}{R^3 B_\text{\tiny QED}}} + \ln(2) - \frac{1}{2} + 12  \zeta^{(1)}(-1) \right ].
\end{split}
\end{equation}

\citet{1997JPhA...30.6475H} also found a similar logarithmic dependence for the radiative corrections to a static magnetic dipole
\footnotesize
\begin{equation}
m_\text{\tiny 1}(\infty) = \alpha_\text{\tiny QED} m_\text{\tiny 0} \left [\frac{1}{3\pi} - \frac{4 \sqrt{3}}{243} \right ] \left [ \ln \left ( \frac{m_\text{\tiny 0}}{R^3 B_\text{\tiny QED}} \right ) -2 \right ] .
\end{equation}
\normalsize
However, the corrections in the case of a rotating dipole are a factor of about two smaller than found in \cite{1997JPhA...30.6475H}.

Given the similarity both physically and mathematically of the two results, we can use the results from \cite{1997JPhA...30.6475H} to provide an interpolation of the effect between the weak and strong-field regimes (see Fig.~\ref{fig:delta_new}). We achieve this by scaling the earlier results both in the magnitude of the effect and the strength of the field to yield the magenta dotted curve depicted in Fig.~\ref{fig:delta_new}. 

\subsection{Discussion}

In order to support our result, we can determine the theoretical strength of the magnetic field at the surface of a magnetar which would lead to
\begin{equation}
\frac{P_\text{\tiny 1,D}(R)}{P_\text{\tiny 0,D}} \underset{R \ll R_\text{\tiny lc}}{\approx} 1.
\end{equation}
We find
\begin{equation}
    B_\text{\tiny p} \approx B_\text{\tiny QED}e^{\frac{3 \pi}{\alpha_\text{\tiny QED}}} \approx 10^{1291}B_\text{\tiny QED}.
\end{equation}
Consequently, for all physically interesting field strengths ($B_\text{\tiny p} \lesssim 4.4 \times 10^{1304}~\text{G}$) the QED radiative corrections to the spin-down are small.

Finally, given the known functional dependence on the magnetic field, in both the weak-field and strong-field limits, of the two-loop corrections, one could wonder whether using the effective Lagrangian of QED to two-loop order would affect our results. According to \citet{2017JHEP...03..108G}, in the weak-field limit, the two-loop Lagrangian is a factor of about $\alpha_\text{\tiny QED}$ smaller than the one-loop Lagrangian, therefore these two-loop corrections do not affect our results in that regime. In the strong-field limit, however, we have \cite{2017JHEP...03..108G}
\begin{equation}
    \frac{\mathcal{L}_\text{\tiny 1}^\text{\tiny 2-loop}}{\mathcal{L}_\text{\tiny 1}^\text{\tiny 1-loop}} \sim \alpha_\text{\tiny QED} \ln(\sqrt{I}).
\end{equation}
Thus, in order for the two-loop corrections to dominate over the one-loop corrections, exponentially large magnetic fields would be needed; such fields are not realized in physically realistic neutron stars. Consequently, we do not expect two-loop corrections to change our conclusions on the importance of QED effects on neutron-star spin-down.

\begin{figure*}
\includegraphics[width=14cm, trim={0 0 0 1cm},clip]{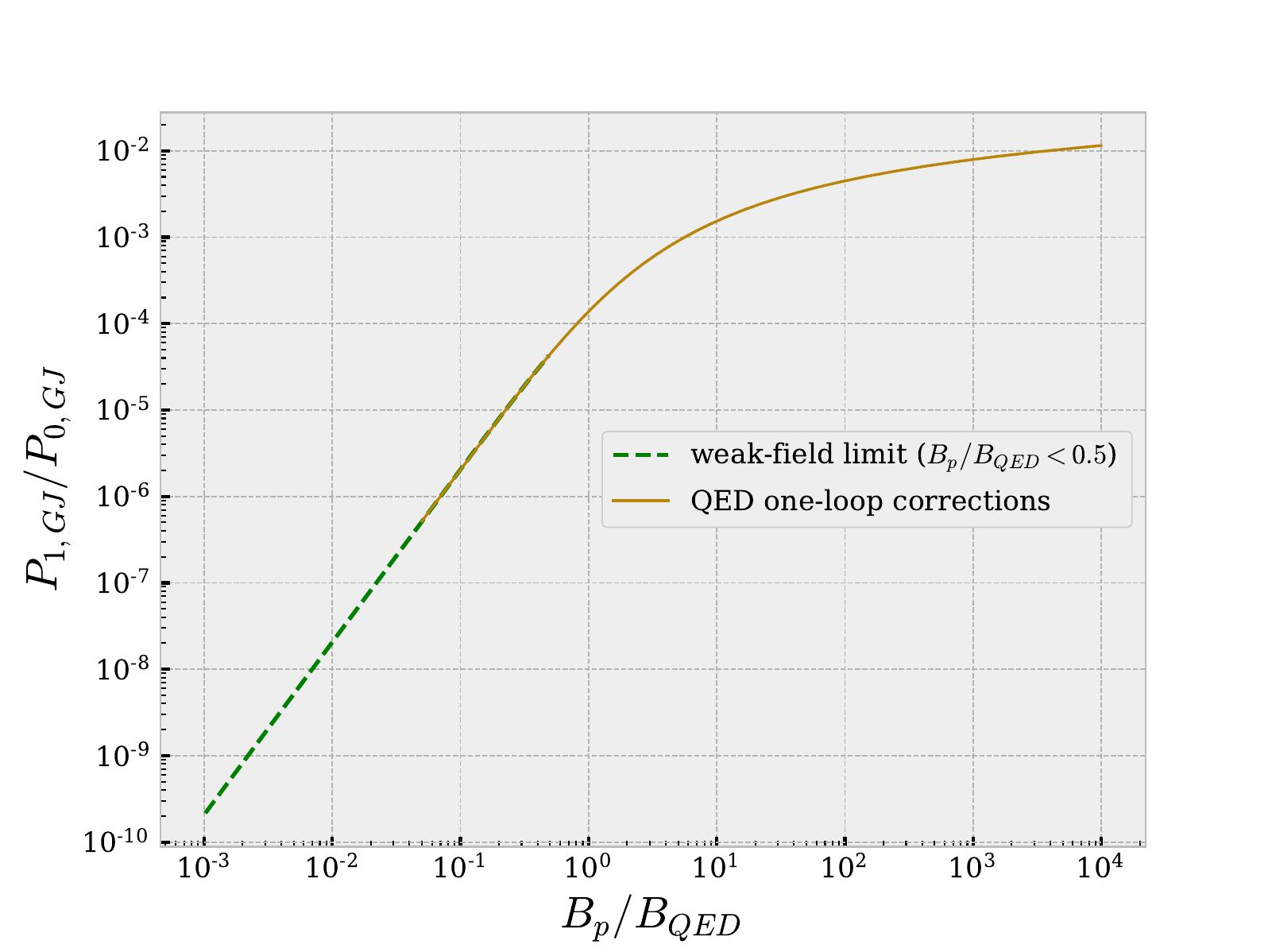}
\caption{The energy loss induced by QED for a Goldreich and Julian pulsar magnetosphere ($P_\text{\tiny 1,GJ}/P_\text{\tiny 0,GJ}$) as a function of the polar magnetic field ($B_\text{\tiny p}$).}
\label{fig:GJ}
\end{figure*}

\section{Energy flow in GJ magnetosphere}
\citet{1969ApJ...157..869G} demonstrated that neutron stars must have a dense corotating magnetosphere within the light cylinder, associated with a wind zone outside the light cylinder. According to the authors, the field has two components, a poloidal one which dominates within the light cylinder, and a toroidal one which dominates within the wind zone. They used an aligned-dipole model for their demonstration. Although an aligned dipole in vacuum does not radiate, the toroidal structure of the magnetic field, in the wind zone, is associated with a non-null Poynting flow, hence a radiated electromagnetic power $P_\text{\tiny 0,GJ}$. In their model, the authors have disregarded both inertia and gravity; thus, the particles in the magnetosphere behave like a perfect conductor, which implies
\begin{equation}
\label{conductor}
    \mathbfit{E} \bm{\cdot} \mathbfit{B} = 0.
\end{equation}

Consequently, the entire flow of angular momentum is carried away by the magnetic field and the total spin-down power over both hemispheres is given by
\begin{equation}
    P_\text{\tiny 0,GJ} = \frac{2\Omega^2}{c} \int_0^{\pi/2} \sin^3 \theta \left [ \Psi (\theta) \right]^2 \, \text{d} \theta,
\end{equation}
where $\Psi(\theta)/r^2$ is the strength of the approximately radial poloidal magnetic field in the wind zone at an angle of $\theta$ relative to the rotation axis.

According to \citet{1969ApJ...157..869G}, the magnetic flux in the asymptotic wind zone can be approximated by the one leaving the polar cap of a neutron star (respectively for each hemisphere). All the field lines emitted inside the polar cap  go through the light cylinder and are open. The bounding field line of the corotating magnetosphere is such that \cite{1969ApJ...157..869G,1971ApJ...164..529S}
\begin{equation}
    \sin\theta_\text{\tiny p} = \left(\frac{\Omega R}{c}\right)^{\frac{1}{2}},
\end{equation}
where $\theta_\text{\tiny p}$ is the polar cap half-angle. The magnetic flux in the asymptotic wind zone is equal to the magnetic flux that leaves the polar cap of the star ($\theta<\theta_\text{\tiny p}$) \cite{1969ApJ...157..869G},
\begin{equation}
     I_\text{\tiny A} = \int_{0}^{\pi/2} \sin\theta \Psi(\theta)  \, \text{d}\theta = \int_{0}^{\theta_\text{\tiny p}} B_\text{\tiny p} R^2 \sin\theta \, \text{d}\theta.
\end{equation} 
Yet, for observed pulsars, we generally have $\theta_\text{\tiny p}\ll 1$, so \cite{1969ApJ...157..869G}
\begin{equation}
     I_\text{\tiny A}\underset{\theta_\text{\tiny p}\ll 1}{=}\frac{1}{2} B_\text{\tiny p} R^2 \theta_\text{\tiny p}^2 = \frac{1}{2}\frac{\Omega R^3}{c} B_\text{\tiny p}.
\end{equation}
We now may write the energy loss in the asymptotic wind zone as follows \cite{1969ApJ...157..869G}
\begin{equation}
    P_\text{\tiny 0,GJ} = \frac{\Omega^2}{c}I_\text{\tiny A}^2 I_\text{\tiny B} =  \frac{1}{4}\frac{\Omega^4 R^6}{c^3} B_\text{\tiny p}^2 I_\text{\tiny B},
\end{equation}
where
\begin{equation}
    I_\text{\tiny B} = \frac{2}{I_\text{\tiny A}^2}\int_{0}^{\pi/2}  \sin^3 \theta \left [ \Psi (\theta) \right]^2 \, \text{d} \theta.
\end{equation}

Since $I_\text{\tiny B}$ takes account of the dispersion of the magnetic flux far away from the light cylinder and relies on geometrical considerations, we do not expect QED to affect this quantity. \citet{1969ApJ...157..869G} assume $I_\text{\tiny B}$ to be of order unity. On the other hand, $I_\text{\tiny A}$ is directly related to the magnetic flux at the polar cap, where the magnetic field is at its strongest; therefore QED should modify this quantity, by increasing the polar magnetic field. 

\subsection{QED One-Loop corrections}
We now consider the QED one-loop corrections to the magnetic field at the polar cap. We treat the general case for the magnetic field strength (the weak-field limit is discussed in the next section). According to Eq.~(\ref{magnetization_field}) and Eq.~(\ref{conductor}), the correction to the magnetic flux leaving the polar cap is the following
\begin{equation}
    I_\text{\tiny A}^\text{\tiny QED} \underset{\theta_\text{\tiny p}\ll 1}{=}\frac{1}{2}B_\text{\tiny p}4\frac{\uppartial \mathcal{L}_\text{\tiny 1}}{\uppartial I} R^2 \theta_\text{\tiny p}^2.
\end{equation}

\citet{1997JPhA...30.6475H}, as well as \citet{1976JETP...42..774R} and \citet{1976JPhA....9.1171D}, derived $\uppartial \mathcal{L}_\text{\tiny 1}/\uppartial I$ as follows
\small
\begin{equation}
    \frac{\uppartial \mathcal{L}_\text{\tiny 1}}{\uppartial I} = \frac{\alpha_\text{\tiny QED}}{8\pi}\left[2X_\text{\tiny 0}\left(\frac{B_\text{\tiny QED}}{B_\text{\tiny p}}\right) - \frac{B_\text{\tiny QED}}{B_\text{\tiny p}}X_\text{\tiny 0}^\text{\tiny (1)}\left(\frac{B_\text{\tiny QED}}{B_\text{\tiny p}}\right)\right],
\end{equation}
\normalsize
where $X_\text{\tiny 0}(x)$ is given by Eq.~(22) of \cite{1997JPhA...30.6475H}, and $X_\text{\tiny 0}^\text{\tiny (1)}(x) = \text{d}X_\text{\tiny 0}(x)/\text{d}x$.

Therefore, the additional QED radiated power is given, to first order in $\alpha_\text{\tiny QED}$, by
\begin{equation}
    P_\text{\tiny 1,GJ} = 2 \frac{\Omega^2}{c} I_\text{\tiny A}I_\text{\tiny A}^\text{\tiny QED}.
\end{equation}

Thus, the amount of energy loss due to QED is the following 
\footnotesize
\begin{equation}
\begin{split}
  \frac{P_\text{\tiny 1,GJ}}{P_\text{\tiny 0,GJ}} =~&  \frac{2\alpha_\text{\tiny QED}}{3 \pi} \left\{12\int_{0}^{\frac{B_\text{\tiny QED}}{2B_\text{\tiny p}}-1} \ln\left[\Gamma(x+1)\right] \, dx \right.\\
  & \left. + \ln\left(\frac{B_\text{\tiny p}}{B_\text{\tiny QED}}\right)+6\ln\pi+7\ln2+12~\zeta^{(1)}(-1) \right. \\
  & \left. -\frac{1}{2}\vphantom{\int_{0}^{\frac{B_\text{\tiny QED}}{2B_\text{\tiny p}}-1} \ln\left[\Gamma(x+1)\right] \, dx}\right\} + \frac{2\alpha_\text{\tiny QED}}{3 \pi} \frac{B_\text{\tiny QED}}{B_\text{\tiny p}}\left\{- 3\ln\left[\Gamma\left(\frac{B_\text{\tiny QED}}{2B_\text{\tiny p}}\right)\right] \right. \\
  &  \left. - \frac{3}{2}\ln\left(\frac{2\pi B_\text{\tiny p}}{B_\text{\tiny QED}}\right)    -3 \vphantom{\ln\left[\Gamma\left(\frac{B_\text{\tiny QED}}{2B_\text{\tiny p}}\right)\right]}\right\} + \frac{\alpha_\text{\tiny QED} B_\text{\tiny QED}^2}{2\pi B_\text{\tiny p}^2},
\end{split}
\end{equation}
\normalsize
where $\Gamma$ is the Gamma function. 

We plot this ratio as a function of the polar magnetic field (see the golden curve depicted in Fig.~\ref{fig:GJ}). Although the energy loss due to QED is about 3 times as big as the one found for an ideal solution in vacuum (the Deutsch fields), the QED corrections to the flow of angular momentum are still small. 

\subsection{QED One-Loop corrections in the weak-field limit}
The expression of $X_\text{\tiny 0}(x)$ that we used becomes difficult to calculate numerically in the weak-field limit. We use instead the expansion given by Eq.~(20) of \cite{1997PhRvD..55.2449H}. This yields an expression for the energy loss to lowest order in the field strength of
\begin{equation}
    \frac{P_\text{\tiny 1,GJ}}{P_\text{\tiny 0,GJ}} = \frac{4\alpha_\text{\tiny QED}}{45\pi} \left(\frac{B_\text{\tiny p}}{B_\text{\tiny QED}}\right)^2.
\end{equation}
This result is around 3 times as big as the one found in Eq.~(\ref{weak_ratio}), for the Deutsch fields. We depict the full weak-field expansion for the energy loss  as a function of the polar magnetic field  by the green dashed curve in Fig.~\ref{fig:GJ}.

\subsection{Discussion}
We included in our calculation a magnetosphere for neutron stars, following the model derived by \citet{1969ApJ...157..869G}. We find that the plasma loading of the magnetosphere of neutron stars yields an energy flow of about the same order as the vacuum result that we obtained with the Deutsch fields. We find that QED effects are also negligible for a pulsar surrounded by a dense magnetosphere (see Fig.~\ref{fig:GJ}). 

We employed a dipole field structure as a first approximation in the Goldreich and Julian model, whereas the plasma influences the field morphology. One could then use a field structure generated by magnetohydrodynamics simulations for an oblique pulsar magnetosphere \cite{2006ApJ...648L..51S}. After measuring the integrated Poynting flux, \citet{2006ApJ...648L..51S} finds the following oblique spin-down luminosity
\begin{equation}
    L_\text{\tiny pulsar} = \frac{\Omega^4 m^2}{c^3}(1+\sin^2\alpha).
\end{equation}
This luminosity is at the minimum $1.5$ times as big as the vacuum formula [see Eq.~(\ref{energy_em_D_appr})]. Again for this more complicated magnetosphere we expect the same geometric arguments that we use for the aligned rotator to apply; therefore, we expect QED to be negligible, even if we use such a field structure.

\section{Conclusions}

Neutron stars are astrophysical objects with strong magnetic fields, especially magnetars for which they can be of order of $B_\text{\tiny p} \approx 10^{15}$~G. Because these objects rotate, they have a rotational energy which serves for the activity of a pulsar. Therefore, a neutron star loses energy and spins down. In the simplest model of a neutron star, a rotating magnetic dipole in rotation in vacuum, the radiated power is given by the classical dipole formula. 

Considering the magnitude of the fields in a neutron star, one could expect quantum electrodynamics to play a role in the energy loss, by a process coined as quantum vacuum friction by \citet{2008EL.....8269002D}. They claimed that a self-torque between a neutron star and the induced magnetization surrounding it will bring its rotation to rest much more quickly that the classical dipole formula would suggest. We demonstrated that the energy loss through QVF is small compared to the power radiated by a rotating magnetic dipole. Then, we calculated the QED one-loop corrections to the Poynting vector, using the local external Deutsch fields of a neutron star. These QED corrections depend on the strength of the magnetic field, so we had to consider two limits, the weak-field limit and the strong-field limit. We obtained, for both of these limits, the ratio of QED radiated power over classical radiated power. In addition, we derived, again in both limits, the one-loop QED corrections to the magnetic moment of a neutron star in vacuum described by the Deutsch fields. We came to the conclusion that, in the weak-field limit as in the strong-field limit (magnetars), the additional radiated power due to QED is small compared to the classical radiated power.

These conclusions do not change for a neutron star surrounded by a dense magnetosphere. Although one could push that study further by using a field structure from magnetohydrodynamics simulations, we expect that this would only introduce additional geometric considerations and therefore we would reach in this most general case the identical conclusion: QED effects on the spin-down luminosity of a neutron star are negligible.

\begin{acknowledgments}
This work was supported by the Natural Sciences and Engineering Research Council of Canada, the Canada Foundation for Innovation, the British Columbia Knowledge Development Fund. This research has made use of NASA's Astrophysics Data System Bibliographic Services and arXiv.org. Finally, we thank the anonymous referees for carefully reading the manuscript and providing valuable comments that improved this manuscript substantially. 
\end{acknowledgments}

\onecolumngrid
\newpage
\appendix*

\section{Vectorial expression of the Deutsch magnetic field}
\label{vectorial_Deutsch_general}
\citet{1999PhR...318..227M} gave the following decomposition of the Deutsch fields
\small
\begin{equation} 
\begin{split}
\mathbfit{H}_\text{\textbf{\tiny D}} =~& \mathbfit{H}_\text{\textbf{\tiny D}}(\text{aligned}) + \mathbfit{H}_\text{\textbf{\tiny D}}(\text{dipole}) + \mathbfit{H}_\text{\textbf{\tiny D}}(\text{quadrupole})\\
\mathbfit{E}_\text{\textbf{\tiny D}} =~& \mathbfit{E}_\text{\textbf{\tiny D}}(\text{aligned}) + \mathbfit{E}_\text{\textbf{\tiny D}}(\text{dipole}) + \mathbfit{E}_\text{\textbf{\tiny D}}(\text{quadrupole})
\label{decomposition}
\end{split}
\end{equation}
According to that decomposition, we have derived the vectorial expression of the Deutsch magnetic field
given, in terms of spherical coordinates, by Eqs. (\ref{Deutsch_magnetic_r}), (\ref{Deutsch_magnetic_theta}) and (\ref{Deutsch_magnetic_phi}) \cite{1955AnAp...18....1D,1999PhR...318..227M}.

\subsection{Aligned part of the Deutsch magnetic field}
We have derived here the vectorial expression of the aligned part of the Deutsch magnetic field

\begin{equation}
\mathbfit{H}_\text{\textbf{\tiny D}}^\text{\textbf{\tiny aligned }}(\mathbfit{r}, t - R/c + r/c) =~ \frac{3}{\Omega^2 r^5} \mathbfit{r} (\mathbfit{m}(t) \bm{\cdot} \bm{\Omega}) (\mathbfit{r} \bm{\cdot} \bm{\Omega}) 
 - \frac{1}{\Omega^2 r^3} \bm{\Omega} (\mathbfit{m}(t) \bm{\cdot} \bm{\Omega})
\end{equation}

\subsection{Dipole part of the Deutsch magnetic field}
We have derived here the vectorial expression of the dipole part of the Deutsch magnetic field
\small
\begin{equation} 
	\begin{split} 
\mathbfit{H}_\text{\textbf{\tiny D}}^\text{\textbf{\tiny dipole}}(\mathbfit{r}, t - R/c + r/c) =~& 
\frac{1}{\Omega^2 R^2 + c^2} \left [
\left ( \frac{1}{r^3} - \frac{3 R}{r^4} 
-~\frac{3 c^2}{\Omega^2 r^5} \right ) \mathbfit{r} (\mathbfit{m}(t) \bm{\cdot} \bm{\Omega})(\mathbfit{r} \bm{\cdot} \bm{\Omega})~+ 
\right.\\
 & \left. 
\left ( \frac{\Omega^2 R}{c r} + \frac{c}{r^2} - \frac{R c}{r^3} \right ) \mathbfit{m}(t) \bm{\times} \bm{\Omega}~+ 
\left ( - \frac{\Omega^2 R}{c r^3} - \frac{3 c}{r^4} + \frac{3 R c}{r^5} \right ) \mathbfit{r}[(\mathbfit{m}(t) \bm{\times} \bm{\Omega}) \bm{\cdot} \mathbfit{r}]~+ 
 \right.\\
 & \left. 
\left ( - \frac{1}{r} + \frac{R}{r^2} + \frac{c^2}{\Omega^2 r^3} \right ) (\mathbfit{m}(t) \bm{\times} \bm{\Omega}) \bm{\times} \bm{\Omega}~+ 
\left ( - \frac{\Omega^2}{r^3} + \frac{3\Omega^2 R}{r^4} + \frac{3 c^2}{r^5} \right ) \mathbfit{r} (\mathbfit{m}(t) \bm{\cdot} \mathbfit{r}) \right ] \\
\end {split}
\end{equation}
\normalsize
\subsection{Quadrupole part of the Deutsch magnetic field}
We have derived here the vectorial expression of the quadrupole part of the Deutsch magnetic field
\small
\begin{equation} 
	\begin{split} 
\mathbfit{H}_\text{\textbf{\tiny D}}^\text{\textbf{\tiny quadrupole}}(\mathbfit{r}, t - R/c + r/c) =~&
\frac{1}{(\Omega^6 R^6 - 3\Omega^4 R^4 c^2 + 36c^2) cr^3} \left \{ \vphantom{\frac{3 c^2}{\Omega^2 r^5}} \right. \\
& \left. \left ( -3\Omega^4 R^4c+6 \Omega^2 R^2 c^3 +\frac{3 \Omega^4 R^5 c-18 \Omega^2 R^3 c^3}{r} + \frac{9 \Omega^2 R^4 c^3-18 R^2 c^5}{r^2}\right ) \mathbfit{r} (\mathbfit{m}(t) \bm{\cdot} \bm{\Omega}) (\mathbfit{r} \bm{\cdot} \bm{\Omega})~+ \right.\\
& \left. \left [ (-\Omega^6 R^5 + 6\Omega^4 R^3 c^2)r^2 + (-9\Omega^4 R^4 c^2 + 18 \Omega^2 R^2 c^4)r + 3 \Omega^4 R^5 c^2-18 \Omega^2 R^3 c^4 \right ] \mathbfit{m}(t) \bm{\times} \bm{\Omega}~+ \right. \\
& \left. \left ( \Omega^6 R^5-6 \Omega^4 R^3 c^2+\frac{9 \Omega^4 R^4 c^2-18 \Omega^2 R^2 c^4}{r}+\frac{-3 \Omega^4 R^5 c^2+18 \Omega^2 R^3 c^4}{r^2} \right ) \mathbfit{r} [(\mathbfit{m}(t) \bm{\times} \bm{\Omega}) \bm{\cdot} \mathbfit{r}]~+ \right.\\
& \left. \left [ (3 \Omega^4 R^4 c-6 \Omega^2 R^2 c^3) r^2+(-3 \Omega^4 R^5 c+18 \Omega^2 R^3 
c^3) r-9 \Omega^2 R^4 c^3+18 R^2 c^5 \right ] (\mathbfit{m}(t) \bm{\times} \bm{\Omega}) \bm{\times} \bm{\Omega} ~+ \right.\\
& \left. \left ( 3 R^4 \Omega^6 c-6 R^2 \Omega^4 c^3+\frac{-3 \Omega^6 R^5 c+18 \Omega^4 R^3 c^3}{r}+\frac{-9 \Omega^4 R^4 c^3+18 \Omega^2 R^2 c^5}{r^2} \right ) \mathbfit{r} (\mathbfit{m}(t) \bm{\cdot} \mathbfit{r})~+ \right.\\
& \left. \left ( -2 R^5 \Omega^4+12 \Omega^2 R^3 c^2+\frac{18 \Omega^2 R^4 c^2+36 R^2 c^4}{r}+\frac{6 \Omega^2 R^5 c^2-36 R^3 c^4}{r^2}\right ) (\mathbfit{r} \bm{\times} \bm{\Omega}) (\mathbfit{m}(t) \bm{\cdot} \bm{\Omega}) (\mathbfit{r} \bm{\cdot} \bm{\Omega})~+ \right.\\
& \left. \left ( -6 \Omega^4 R^4 c +12 \Omega^2 R^2 c^3 +\frac{6 \Omega^4 R^5 c-36 \Omega^2 R^3 c^3}{r}+\frac{18 \Omega^2 R^4 c^3-36 R^2 c^5}{r^2} \right ) (\mathbfit{r} \bm{\times} \bm{\Omega}) [(\mathbfit{m}(t) \bm{\times} \bm{\Omega}) \bm{\cdot} \mathbfit{r}]~+ \right.\\
& \left. \left ( 2 \Omega^6 R^5-12 \Omega^4 R^3 c^2+ \frac{18 \Omega^4 R^4 c^2-36 \Omega^2 R^2 c^4}{r} + \frac{-6 \Omega^4 R^5 c^2+36 \Omega^2 R^3 c^4}{r^2} \right ) (\mathbfit{r} \bm{\times} \bm{\Omega}) (\mathbfit{m}(t) \bm{\cdot} \mathbfit{r}) \right \}
\end {split}
\end{equation}
\normalsize

\newpage
\twocolumngrid


\begin{thebibliography}{31}%
\makeatletter
\providecommand \@ifxundefined [1]{%
 \@ifx{#1\undefined}
}%
\providecommand \@ifnum [1]{%
 \ifnum #1\expandafter \@firstoftwo
 \else \expandafter \@secondoftwo
 \fi
}%
\providecommand \@ifx [1]{%
 \ifx #1\expandafter \@firstoftwo
 \else \expandafter \@secondoftwo
 \fi
}%
\providecommand \natexlab [1]{#1}%
\providecommand \enquote  [1]{``#1''}%
\providecommand \bibnamefont  [1]{#1}%
\providecommand \bibfnamefont [1]{#1}%
\providecommand \citenamefont [1]{#1}%
\providecommand \href@noop [0]{\@secondoftwo}%
\providecommand \href [0]{\begingroup \@sanitize@url \@href}%
\providecommand \@href[1]{\@@startlink{#1}\@@href}%
\providecommand \@@href[1]{\endgroup#1\@@endlink}%
\providecommand \@sanitize@url [0]{\catcode `\\12\catcode `\$12\catcode
  `\&12\catcode `\#12\catcode `\^12\catcode `\_12\catcode `\%12\relax}%
\providecommand \@@startlink[1]{}%
\providecommand \@@endlink[0]{}%
\providecommand \url  [0]{\begingroup\@sanitize@url \@url }%
\providecommand \@url [1]{\endgroup\@href {#1}{\urlprefix }}%
\providecommand \urlprefix  [0]{URL }%
\providecommand \Eprint [0]{\href }%
\providecommand \doibase [0]{http://dx.doi.org/}%
\providecommand \selectlanguage [0]{\@gobble}%
\providecommand \bibinfo  [0]{\@secondoftwo}%
\providecommand \bibfield  [0]{\@secondoftwo}%
\providecommand \translation [1]{[#1]}%
\providecommand \BibitemOpen [0]{}%
\providecommand \bibitemStop [0]{}%
\providecommand \bibitemNoStop [0]{.\EOS\space}%
\providecommand \EOS [0]{\spacefactor3000\relax}%
\providecommand \BibitemShut  [1]{\csname bibitem#1\endcsname}%
\let\auto@bib@innerbib\@empty
\bibitem [{\citenamefont {{Dupays}}\ \emph {et~al.}(2008)\citenamefont
  {{Dupays}}, \citenamefont {{Rizzo}}, \citenamefont {{Bakalov}},\ and\
  \citenamefont {{Bignami}}}]{2008EL.....8269002D}%
  \BibitemOpen
  \bibfield  {author} {\bibinfo {author} {\bibfnamefont {A.}~\bibnamefont
  {{Dupays}}}, \bibinfo {author} {\bibfnamefont {C.}~\bibnamefont {{Rizzo}}},
  \bibinfo {author} {\bibfnamefont {D.}~\bibnamefont {{Bakalov}}}, \ and\
  \bibinfo {author} {\bibfnamefont {G.~F.}\ \bibnamefont {{Bignami}}},\ }\href
  {\doibase 10.1209/0295-5075/82/69002} {\bibfield  {journal} {\bibinfo
  {journal} {EPL (Europhysics Letters)}\ }\textbf {\bibinfo {volume} {82}},\
  \bibinfo {pages} {69002} (\bibinfo {year} {2008})}\BibitemShut {NoStop}%
\bibitem [{\citenamefont {{Coelho}}\ \emph {et~al.}(2016)\citenamefont
  {{Coelho}}, \citenamefont {{Pereira}},\ and\ \citenamefont {{de
  Araujo}}}]{2016ApJ...823...97C}%
  \BibitemOpen
  \bibfield  {author} {\bibinfo {author} {\bibfnamefont {J.~G.}\ \bibnamefont
  {{Coelho}}}, \bibinfo {author} {\bibfnamefont {J.~P.}\ \bibnamefont
  {{Pereira}}}, \ and\ \bibinfo {author} {\bibfnamefont {J.~C.~N.}\
  \bibnamefont {{de Araujo}}},\ }\href {\doibase 10.3847/0004-637X/823/2/97}
  {\bibfield  {journal} {\bibinfo  {journal} {\apj}\ }\textbf {\bibinfo
  {volume} {823}},\ \bibinfo {eid} {97} (\bibinfo {year} {2016})}\BibitemShut
  {NoStop}%
\bibitem [{\citenamefont {{Xiong}}\ \emph {et~al.}(2016)\citenamefont
  {{Xiong}}, \citenamefont {{Gao}},\ and\ \citenamefont
  {{Xu}}}]{2016RAA....16....9X}%
  \BibitemOpen
  \bibfield  {author} {\bibinfo {author} {\bibfnamefont {X.-Y.}\ \bibnamefont
  {{Xiong}}}, \bibinfo {author} {\bibfnamefont {C.-Y.}\ \bibnamefont {{Gao}}},
  \ and\ \bibinfo {author} {\bibfnamefont {R.-X.}\ \bibnamefont {{Xu}}},\
  }\href {\doibase 10.1088/1674-4527/16/1/009} {\bibfield  {journal} {\bibinfo
  {journal} {Research in Astronomy and Astrophysics}\ }\textbf {\bibinfo
  {volume} {16}},\ \bibinfo {eid} {9} (\bibinfo {year} {2016})}\BibitemShut
  {NoStop}%
\bibitem [{\citenamefont {{Dupays}}\ \emph {et~al.}(2012)\citenamefont
  {{Dupays}}, \citenamefont {{Rizzo}},\ and\ \citenamefont {{Fabrizio
  Bignami}}}]{2012EL.....9849001D}%
  \BibitemOpen
  \bibfield  {author} {\bibinfo {author} {\bibfnamefont {A.}~\bibnamefont
  {{Dupays}}}, \bibinfo {author} {\bibfnamefont {C.}~\bibnamefont {{Rizzo}}}, \
  and\ \bibinfo {author} {\bibfnamefont {G.}~\bibnamefont {{Fabrizio
  Bignami}}},\ }\href {\doibase 10.1209/0295-5075/98/49001} {\bibfield
  {journal} {\bibinfo  {journal} {EPL (Europhysics Letters)}\ }\textbf
  {\bibinfo {volume} {98}},\ \bibinfo {pages} {49001} (\bibinfo {year}
  {2012})}\BibitemShut {NoStop}%
\bibitem [{\citenamefont {{Abraham}}(1902)}]{1902AnP...315..105A}%
  \BibitemOpen
  \bibfield  {author} {\bibinfo {author} {\bibfnamefont {M.}~\bibnamefont
  {{Abraham}}},\ }\href {\doibase 10.1002/andp.19023150105} {\bibfield
  {journal} {\bibinfo  {journal} {Annalen der Physik}\ }\textbf {\bibinfo
  {volume} {315}},\ \bibinfo {pages} {105} (\bibinfo {year}
  {1902})}\BibitemShut {NoStop}%
\bibitem [{\citenamefont {{Abraham}}(1904)}]{1904AnP...319..236A}%
  \BibitemOpen
  \bibfield  {author} {\bibinfo {author} {\bibfnamefont {M.}~\bibnamefont
  {{Abraham}}},\ }\href {\doibase 10.1002/andp.19043190703} {\bibfield
  {journal} {\bibinfo  {journal} {Annalen der Physik}\ }\textbf {\bibinfo
  {volume} {319}},\ \bibinfo {pages} {236} (\bibinfo {year}
  {1904})}\BibitemShut {NoStop}%
\bibitem [{\citenamefont {Harte}(2006)}]{PhysRevD.73.065006}%
  \BibitemOpen
  \bibfield  {author} {\bibinfo {author} {\bibfnamefont {A.~I.}\ \bibnamefont
  {Harte}},\ }\href {\doibase 10.1103/PhysRevD.73.065006} {\bibfield  {journal}
  {\bibinfo  {journal} {Phys. Rev. D}\ }\textbf {\bibinfo {volume} {73}},\
  \bibinfo {pages} {065006} (\bibinfo {year} {2006})}\BibitemShut {NoStop}%
\bibitem [{\citenamefont {{Bonga}}\ \emph {et~al.}(2018)\citenamefont
  {{Bonga}}, \citenamefont {{Poisson}},\ and\ \citenamefont
  {{Yang}}}]{2018AmJPh..86..839B}%
  \BibitemOpen
  \bibfield  {author} {\bibinfo {author} {\bibfnamefont {B.}~\bibnamefont
  {{Bonga}}}, \bibinfo {author} {\bibfnamefont {E.}~\bibnamefont {{Poisson}}},
  \ and\ \bibinfo {author} {\bibfnamefont {H.}~\bibnamefont {{Yang}}},\ }\href
  {\doibase 10.1119/1.5054590} {\bibfield  {journal} {\bibinfo  {journal} {Am.
  J. Phys.}\ }\textbf {\bibinfo {volume} {86}},\ \bibinfo {pages} {839}
  (\bibinfo {year} {2018})}\BibitemShut {NoStop}%
\bibitem [{\citenamefont {{Gralla}}\ \emph {et~al.}(2009)\citenamefont
  {{Gralla}}, \citenamefont {{Harte}},\ and\ \citenamefont
  {{Wald}}}]{2009PhRvD..80b4031G}%
  \BibitemOpen
  \bibfield  {author} {\bibinfo {author} {\bibfnamefont {S.~E.}\ \bibnamefont
  {{Gralla}}}, \bibinfo {author} {\bibfnamefont {A.~I.}\ \bibnamefont
  {{Harte}}}, \ and\ \bibinfo {author} {\bibfnamefont {R.~M.}\ \bibnamefont
  {{Wald}}},\ }\href {\doibase 10.1103/PhysRevD.80.024031} {\bibfield
  {journal} {\bibinfo  {journal} {\prd}\ }\textbf {\bibinfo {volume} {80}},\
  \bibinfo {eid} {024031} (\bibinfo {year} {2009})}\BibitemShut {NoStop}%
\bibitem [{\citenamefont {{Deutsch}}(1955)}]{1955AnAp...18....1D}%
  \BibitemOpen
  \bibfield  {author} {\bibinfo {author} {\bibfnamefont {A.~J.}\ \bibnamefont
  {{Deutsch}}},\ } {\bibfield  {journal} {\bibinfo  {journal}
  {Annales d'Astrophysique}\ }\textbf {\bibinfo {volume} {18}},\ \bibinfo
  {pages} {1} (\bibinfo {year} {1955})}\BibitemShut {NoStop}%
\bibitem [{\citenamefont {{Heyl}}\ and\ \citenamefont
  {{Hernquist}}(1997{\natexlab{a}})}]{1997JPhA...30.6485H}%
  \BibitemOpen
  \bibfield  {author} {\bibinfo {author} {\bibfnamefont {J.~S.}\ \bibnamefont
  {{Heyl}}}\ and\ \bibinfo {author} {\bibfnamefont {L.}~\bibnamefont
  {{Hernquist}}},\ }\href {\doibase 10.1088/0305-4470/30/18/022} {\bibfield
  {journal} {\bibinfo  {journal} {J. Phys. A}\ }\textbf {\bibinfo {volume}
  {30}},\ \bibinfo {pages} {6485} (\bibinfo {year}
  {1997}{\natexlab{a}})}\BibitemShut {NoStop}%
\bibitem [{\citenamefont {{Pétri}}(2016{\natexlab{a}})}]{2016A&A...594A.112P}%
  \BibitemOpen
  \bibfield  {author} {\bibinfo {author} {\bibfnamefont {J.}~\bibnamefont
  {{Pétri}}},\ }\href {\doibase 10.1051/0004-6361/201628518} {\bibfield
  {journal} {\bibinfo  {journal} {Astron. Astrophys.}\ }\textbf {\bibinfo
  {volume} {594}},\ \bibinfo {eid} {A112} (\bibinfo {year}
  {2016}{\natexlab{a}})}\BibitemShut {NoStop}%
\bibitem [{\citenamefont {{Pétri}}(2016{\natexlab{b}})}]{2016MNRAS.456.4455P}%
  \BibitemOpen
  \bibfield  {author} {\bibinfo {author} {\bibfnamefont {J.}~\bibnamefont
  {{Pétri}}},\ }\href {\doibase 10.1093/mnras/stv2967} {\bibfield
  {journal} {\bibinfo  {journal} {Mon. Not. R. Astron. Soc.}\ }\textbf
  {\bibinfo {volume} {456}},\ \bibinfo {pages} {4455} (\bibinfo {year}
  {2016}{\natexlab{b}})}\BibitemShut {NoStop}%
\bibitem [{\citenamefont {{Goldreich}}\ and\ \citenamefont
  {{Julian}}(1969)}]{1969ApJ...157..869G}%
  \BibitemOpen
  \bibfield  {author} {\bibinfo {author} {\bibfnamefont {P.}~\bibnamefont
  {{Goldreich}}}\ and\ \bibinfo {author} {\bibfnamefont {W.~H.}\ \bibnamefont
  {{Julian}}},\ }\href {\doibase 10.1086/150119} {\bibfield  {journal}
  {\bibinfo  {journal} {\apj}\ }\textbf {\bibinfo {volume} {157}},\ \bibinfo
  {pages} {869} (\bibinfo {year} {1969})}\BibitemShut {NoStop}%
\bibitem [{\citenamefont {{Jackson}}(1998)}]{1998clel.book.....J}%
  \BibitemOpen
  \bibfield  {author} {\bibinfo {author} {\bibfnamefont {J.~D.}\ \bibnamefont
  {{Jackson}}},\ }\href
  {https://www.wiley.com/en-us/Classical+Electrodynamics\%2C+3rd+Edition-p-9780471309321}
  {\emph {\bibinfo {title} {Classical Electrodynamics}}},\ \bibinfo {edition}
  {3rd}\ ed.\ (\bibinfo  {publisher} {Wiley-VCH},\ \bibinfo {year}
  {1998})\BibitemShut {NoStop}%
\bibitem [{\citenamefont {{Kinsler}}\ \emph {et~al.}(2009)\citenamefont
  {{Kinsler}}, \citenamefont {{Favaro}},\ and\ \citenamefont
  {{McCall}}}]{2009EJPh...30..983K}%
  \BibitemOpen
  \bibfield  {author} {\bibinfo {author} {\bibfnamefont {P.}~\bibnamefont
  {{Kinsler}}}, \bibinfo {author} {\bibfnamefont {A.}~\bibnamefont {{Favaro}}},
  \ and\ \bibinfo {author} {\bibfnamefont {M.~W.}\ \bibnamefont {{McCall}}},\
  }\href {\doibase 10.1088/0143-0807/30/5/007} {\bibfield  {journal} {\bibinfo
  {journal} {European Journal of Physics}\ }\textbf {\bibinfo {volume} {30}},\
  \bibinfo {pages} {983} (\bibinfo {year} {2009})}\BibitemShut {NoStop}%
\bibitem [{\citenamefont {{Heisenberg}}\ and\ \citenamefont
  {{Euler}}(1936)}]{1936ZPhy...98..714H}%
  \BibitemOpen
  \bibfield  {author} {\bibinfo {author} {\bibfnamefont {W.}~\bibnamefont
  {{Heisenberg}}}\ and\ \bibinfo {author} {\bibfnamefont {H.}~\bibnamefont
  {{Euler}}},\ }\href {\doibase 10.1007/BF01343663} {\bibfield  {journal}
  {\bibinfo  {journal} {Zeitschrift fur Physik}\ }\textbf {\bibinfo {volume}
  {98}},\ \bibinfo {pages} {714} (\bibinfo {year} {1936})}\BibitemShut
  {NoStop}%
\bibitem [{\citenamefont {{Heyl}}\ and\ \citenamefont
  {{Hernquist}}(1997{\natexlab{b}})}]{1997PhRvD..55.2449H}%
  \BibitemOpen
  \bibfield  {author} {\bibinfo {author} {\bibfnamefont {J.~S.}\ \bibnamefont
  {{Heyl}}}\ and\ \bibinfo {author} {\bibfnamefont {L.}~\bibnamefont
  {{Hernquist}}},\ }\href {\doibase 10.1103/PhysRevD.55.2449} {\bibfield
  {journal} {\bibinfo  {journal} {Phys. Rev. D}\ }\textbf {\bibinfo {volume}
  {55}},\ \bibinfo {pages} {2449} (\bibinfo {year}
  {1997}{\natexlab{b}})}\BibitemShut {NoStop}%
\bibitem [{\citenamefont {{Schwinger}}(1951)}]{1951PhRv...82..664S}%
  \BibitemOpen
  \bibfield  {author} {\bibinfo {author} {\bibfnamefont {J.}~\bibnamefont
  {{Schwinger}}},\ }\href {\doibase 10.1103/PhysRev.82.664} {\bibfield
  {journal} {\bibinfo  {journal} {Physical Review}\ }\textbf {\bibinfo {volume}
  {82}},\ \bibinfo {pages} {664} (\bibinfo {year} {1951})}\BibitemShut
  {NoStop}%
\bibitem [{\citenamefont {{Karbstein}}\ and\ \citenamefont
  {{Shaisultanov}}(2015)}]{2015PhRvD..91h5027K}%
  \BibitemOpen
  \bibfield  {author} {\bibinfo {author} {\bibfnamefont {F.}~\bibnamefont
  {{Karbstein}}}\ and\ \bibinfo {author} {\bibfnamefont {R.}~\bibnamefont
  {{Shaisultanov}}},\ }\href {\doibase 10.1103/PhysRevD.91.085027} {\bibfield
  {journal} {\bibinfo  {journal} {\prd}\ }\textbf {\bibinfo {volume} {91}},\
  \bibinfo {eid} {085027} (\bibinfo {year} {2015})}\BibitemShut {NoStop}%
\bibitem [{\citenamefont {{Berestetskii}}\ \emph {et~al.}(2012)\citenamefont
  {{Berestetskii}}, \citenamefont {{Lifshitz}},\ and\ \citenamefont
  {{Pitaevskii}}}]{Berestetskii}%
  \BibitemOpen
  \bibfield  {author} {\bibinfo {author} {\bibfnamefont {V.~B.}\ \bibnamefont
  {{Berestetskii}}}, \bibinfo {author} {\bibfnamefont {E.~M.}\ \bibnamefont
  {{Lifshitz}}}, \ and\ \bibinfo {author} {\bibfnamefont {L.~P.}\ \bibnamefont
  {{Pitaevskii}}},\ }\href
  {https://www.elsevier.com/books/quantum-electrodynamics/berestetskii/978-0-08-050346-2}
  {\emph {\bibinfo {title} {Quantum Electrodynamics}}},\ \bibinfo {edition}
  {2nd}\ ed.,\ Vol.~\bibinfo {volume} {4}\ (\bibinfo  {publisher}
  {Butterworth-Heinemann},\ \bibinfo {year} {2012})\BibitemShut {NoStop}%
\bibitem [{\citenamefont {{Denisov}}\ \emph {et~al.}(2016)\citenamefont
  {{Denisov}}, \citenamefont {{Shvilkin}}, \citenamefont {{Sokolov}},\ and\
  \citenamefont {{Vasili'ev}}}]{2016PhRvD..94d5021D}%
  \BibitemOpen
  \bibfield  {author} {\bibinfo {author} {\bibfnamefont {V.~I.}\ \bibnamefont
  {{Denisov}}}, \bibinfo {author} {\bibfnamefont {B.~N.}\ \bibnamefont
  {{Shvilkin}}}, \bibinfo {author} {\bibfnamefont {V.~A.}\ \bibnamefont
  {{Sokolov}}}, \ and\ \bibinfo {author} {\bibfnamefont {M.~I.}\ \bibnamefont
  {{Vasili'ev}}},\ }\href {\doibase 10.1103/PhysRevD.94.045021} {\bibfield
  {journal} {\bibinfo  {journal} {\prd}\ }\textbf {\bibinfo {volume} {94}},\
  \bibinfo {eid} {045021} (\bibinfo {year} {2016})}\BibitemShut {NoStop}%
\bibitem [{\citenamefont {{Heyl}}\ and\ \citenamefont
  {{Hernquist}}(1997{\natexlab{c}})}]{1997JPhA...30.6475H}%
  \BibitemOpen
  \bibfield  {author} {\bibinfo {author} {\bibfnamefont {J.~S.}\ \bibnamefont
  {{Heyl}}}\ and\ \bibinfo {author} {\bibfnamefont {L.}~\bibnamefont
  {{Hernquist}}},\ }\href {\doibase 10.1088/0305-4470/30/18/021} {\bibfield
  {journal} {\bibinfo  {journal} {J. Phys. A}\ }\textbf {\bibinfo {volume}
  {30}},\ \bibinfo {pages} {6475} (\bibinfo {year}
  {1997}{\natexlab{c}})}\BibitemShut {NoStop}%
\bibitem [{\citenamefont {Greiner}(1998)}]{Greiner}%
  \BibitemOpen
  \bibfield  {author} {\bibinfo {author} {\bibfnamefont {W.}~\bibnamefont
  {Greiner}},\ }\href {\doibase 10.1007/978-1-4612-0587-6} {\emph {\bibinfo
  {title} {Classical Electrodynamics}}},\ \bibinfo {edition} {1st}\ ed.\
  (\bibinfo  {publisher} {Springer-Verlag New York},\ \bibinfo {year}
  {1998})\BibitemShut {NoStop}%
\bibitem [{\citenamefont {{Stump}}\ and\ \citenamefont
  {{Pollack}}(1997)}]{1997AmJPh..65...81S}%
  \BibitemOpen
  \bibfield  {author} {\bibinfo {author} {\bibfnamefont {D.~R.}\ \bibnamefont
  {{Stump}}}\ and\ \bibinfo {author} {\bibfnamefont {G.~L.}\ \bibnamefont
  {{Pollack}}},\ }\href {\doibase 10.1119/1.18523} {\bibfield  {journal}
  {\bibinfo  {journal} {Am. J. Phys.}\ }\textbf {\bibinfo {volume} {65}},\
  \bibinfo {pages} {81} (\bibinfo {year} {1997})}\BibitemShut {NoStop}%
\bibitem [{\citenamefont {{Michel}}\ and\ \citenamefont
  {{Li}}(1999)}]{1999PhR...318..227M}%
  \BibitemOpen
  \bibfield  {author} {\bibinfo {author} {\bibfnamefont {F.~C.}\ \bibnamefont
  {{Michel}}}\ and\ \bibinfo {author} {\bibfnamefont {H.}~\bibnamefont
  {{Li}}},\ }\href {\doibase 10.1016/S0370-1573(99)00002-2} {\bibfield
  {journal} {\bibinfo  {journal} {Physics Reports}\ }\textbf {\bibinfo {volume}
  {318}},\ \bibinfo {pages} {227} (\bibinfo {year} {1999})}\BibitemShut
  {NoStop}%
\bibitem [{\citenamefont {{Ritus}}(1976)}]{1976JETP...42..774R}%
  \BibitemOpen
  \bibfield  {author} {\bibinfo {author} {\bibfnamefont {V.~I.}\ \bibnamefont
  {{Ritus}}},\ } {\bibfield  {journal} {\bibinfo  {journal}
  {Soviet Journal of Experimental and Theoretical Physics}\ }\textbf {\bibinfo
  {volume} {42}},\ \bibinfo {pages} {774} (\bibinfo {year} {1976})}\BibitemShut
  {NoStop}%
\bibitem [{\citenamefont {{Dittrich}}(1976)}]{1976JPhA....9.1171D}%
  \BibitemOpen
  \bibfield  {author} {\bibinfo {author} {\bibfnamefont {W.}~\bibnamefont
  {{Dittrich}}},\ }\href {\doibase 10.1088/0305-4470/9/7/019} {\bibfield
  {journal} {\bibinfo  {journal} {J. Phys. A}\ }\textbf {\bibinfo {volume}
  {9}},\ \bibinfo {pages} {1171} (\bibinfo {year} {1976})}\BibitemShut
  {NoStop}%
\bibitem [{\citenamefont {{Gies}}\ and\ \citenamefont
  {{Karbstein}}(2017)}]{2017JHEP...03..108G}%
  \BibitemOpen
  \bibfield  {author} {\bibinfo {author} {\bibfnamefont {H.}~\bibnamefont
  {{Gies}}}\ and\ \bibinfo {author} {\bibfnamefont {F.}~\bibnamefont
  {{Karbstein}}},\ }\href {\doibase 10.1007/JHEP03(2017)108} {\bibfield
  {journal} {\bibinfo  {journal} {J. High Energy Phys.}\ }\textbf {\bibinfo
  {volume} {2017}},\ \bibinfo {eid} {108} (\bibinfo {year} {2017})}\BibitemShut
  {NoStop}%
\bibitem [{\citenamefont {{Sturrock}}(1971)}]{1971ApJ...164..529S}%
  \BibitemOpen
  \bibfield  {author} {\bibinfo {author} {\bibfnamefont {P.~A.}\ \bibnamefont
  {{Sturrock}}},\ }\href {\doibase 10.1086/150865} {\bibfield  {journal}
  {\bibinfo  {journal} {\apj}\ }\textbf {\bibinfo {volume} {164}},\ \bibinfo
  {pages} {529} (\bibinfo {year} {1971})}\BibitemShut {NoStop}%
\bibitem [{\citenamefont {{Spitkovsky}}(2006)}]{2006ApJ...648L..51S}%
  \BibitemOpen
  \bibfield  {author} {\bibinfo {author} {\bibfnamefont {A.}~\bibnamefont
  {{Spitkovsky}}},\ }\href {\doibase 10.1086/507518} {\bibfield  {journal}
  {\bibinfo  {journal} {\apj}\ }\textbf {\bibinfo {volume} {648}},\ \bibinfo
  {pages} {L51} (\bibinfo {year} {2006})}\BibitemShut {NoStop}%
\end{thebibliography}

%

\end{document}